\documentclass[10pt]{wlscirep}
\usepackage[utf8]{inputenc}
\usepackage[T1]{fontenc}
\usepackage{xcolor}
\usepackage{ragged2e}
\usepackage{cite}
\usepackage{amsmath,amssymb,amsfonts}
\usepackage{algorithm}
\usepackage{graphics}
\usepackage{graphicx}
\usepackage{textcomp}
\usepackage{caption}
\usepackage{subcaption}
\usepackage{colortbl}
\usepackage{color}
\usepackage{hhline}

\definecolor{REVISE_COLOR_1}{RGB}{0,0,0}
\definecolor{REVISE_COLOR_2}{RGB}{0,0,0}

\usepackage{makecell}
\usepackage{algpseudocode}
\usepackage{multirow}
\usepackage[mathlines]{lineno}
\usepackage{float}

\usepackage{nccmath,lipsum}

\title{Enhancing the Yield of Bucket Brigade Quantum Random Access Memory using Redundancy Repair}

\author[1]{Dongmin Kim}
\author[1]{Sengthai Heng}
\author[1]{Sanghyeon Lee}
\author[1,*]{Youngsun Han}
\affil[1]{Department of AI Convergence, Pukyong National University, Nam-gu, Busan 48513, South Korea}

\affil[*]{youngsun@pknu.ac.kr}


\begin{abstract}
Quantum Random Access Memory (qRAM) is an essential computing element for running oracle-based quantum algorithms. qRAM exploits quantum superposition to access all data stored in the memory cells simultaneously and guarantees the superior performance of quantum algorithms.
A qRAM memory cell comprises logical qubits encoded through quantum error correction technology for successful operation against various quantum noises. In addition to quantum noise, the low-technology nodes based on silicon technology can increase the qubit density and may introduce defective qubits. As qRAM comprises many qubits, its yield will be reduced by defective qubits; these qubits must be handled using QEC scheme.
However, the QEC scheme requires numerous physical qubits, which burdens resource overhead. 
In this paper, to resolve this overhead problem, we propose a novel quantum memory architecture that compensates for defective qubits by introducing redundant qubits. 
We also analyze the yield improvement offered by our proposed quantum memory architecture by varying the ideal fabrication error rate from 0.5\% to 1\% for different numbers of logical qubits in the qRAM.
We demonstrate that for the qRAM comprising 1,024 logical qubits, eight redundant logical qubits improved the yield by 95.92\% from that of qRAM not employing the redundant repair scheme.
\end{abstract}
\begin{document}

\flushbottom
\maketitle
\thispagestyle{empty}

\section*{Introduction}

Quantum algorithms promise to solve specific problems that cannot be solved by classical algorithms within a reasonable time. Representative quantum algorithms are the Shor algorithm~\cite{shor1994af} and the Grover algorithm~\cite{grover1996af}. The Shor algorithm divides a huge composite number into two prime factors, and the Grover algorithm finds the desired data among a large dataset. Exploiting the ‘‘superposition’’ phenomenon of qubits, both algorithms simultaneously process all data combinations that can be represented by the qubits, thereby speeding up the calculations. Various quantum algorithms that utilize the advantages of qubit-based computation are currently being developed.

To guarantee that quantum superposition can exponentially increase the speed of quantum algorithms over their classical counterparts, we must load the superposed data into a quantum processor. Most of the representative quantum algorithms used in various fields assume that data can be loaded through an oracle, commonly referred to as a black box~\cite{hann2021po}. Such a black box must be fully supported with some degree of abstraction for oracle-based quantum algorithms. As a quantum oracle, Quantum Random Access Memory (qRAM)~\cite{hann2021po, bang2019ou, giovannetti2008qr} is required for achieving an exponential computational advantage of quantum algorithms over conventional algorithms~\cite {bang2019ou, giovannetti2008af, wang_fundamental_2024, Duan2024ca}. For this purpose, the qRAM must store data in a superposition state. qRAM already guarantees exponential speed improvement in tasks such as data processing~\cite{de2020cb, zidan2021an} and pattern recognition~\cite{soni2021qb, trugenberger2001pq, schaller2006qa, schutzhold2003pr} and is required for other quantum algorithms such as quantum searching~\cite{zajac2022tq, nielsen2010qc}, collision finding~\cite{park2019cb, childs2007wf}, and element distinctness problems~\cite{buhrman2001qa}.

Promising qRAM architectures are broadly divided into Fanout and Bucket Brigade schemes~\cite{arunachalam2015ot, Xu2023sa, hann2021ro}, which adopt trapped ion qubits and qutrits (storage units of three-state information $|wait\rangle$, $|left\rangle$, and $|right\rangle$) as the basic unit, respectively. The two schemes differ mainly by the number of gates activated in their memory-addressing processes. For a single $n$-bit query string, Fanout qRAM activates $O(2^n)$ gates, whereas in Bucket Brigade qRAM activates $O(poly(n))$ gates~\cite{arunachalam2015ot, giovannetti2008af}. In the actual implementation of qRAM, gate activation is considered as physical transistor activation~\cite{arunachalam2015ot,giovannetti2008af}. However, as numerous transistor activations cannot be operated within a relatively short quantum coherence time, they deteriorate the accuracy of the results. Accordingly, the Fanout scheme is more susceptible to decoherence error than the Bucket Brigade scheme.

Various studies have shown that memory addressing is more error-resistant in the Bucket Brigade qRAM scheme than in the Fanout scheme~\cite{arunachalam2015ot, Weiss2024qr}. Bucket Brigade also accelerates memory addressing by parallelizing the queries~\cite{paler2020pt}. These studies focus on the robustness of Bucket Brigade qRAM with a small number of gate activations required for memory addressing. However, to improve the fault tolerance of qRAM, one must consider the robustness of the memory cell qubits comprising the qRAM~\cite{chen2021sa, hann2021po}. Although the Bucket Brigade achieves more error-resistant memory addressing than the Fanout structure, if the memory cells are vulnerable to errors, this memory addressing becomes meaningless.
Despite this necessity, existing studies focusing on fault-tolerant addressing often overlook the integrity of the memory cells themselves. For instance, Weiss et al.~\cite{weiss_faulty_2024} proposed the Faulty Towers architecture. While this approach effectively mitigates manufacturing defects within the routing network to ensure reliable addressing paths, it does not consider scenarios where errors occur in the memory cells serving as the final destination.
Therefore, for fully fault-tolerant qRAM, the memory cell qubits should be error-resistant as well. This requires a quantum error correction (QEC) scheme, which typically spreads data stored in a physical qubit across multiple highly entangled physical qubits. The QEC code scheme creates a ``logical qubit'', a virtual qubit holding the data used in the quantum algorithm~\cite{google2023sq, abobeih2022ft, hua2021aa, molavi2022qm}.

QEC applied to a logical qubit protects the quantum state from general quantum noise. Moreover, QEC must handle defective physical qubits that may occur during the fabrication process. This is because the density of qubits has increased based on recent silicon technology, and as mass production of physical qubits has become possible~\cite{rotta2017qi, zwerver2022qm}. It makes the probability of the occurrence of defective qubits also increase, so that many defective qRAMs are manufactured. As defective qRAMs lead to poor yield, reducing the number of defective logical qubits is essential for improving the qRAM yield. To this end, the QEC for encoding a logical qubit should cover a large number of physical qubits~\cite{smith2022ss, maurya2022cc, heeres2017ia, luo2021qt}. However, increasing the number of required logical qubits in qRAM increases the resource overhead because many additional physical qubits are required to support a higher degree of QEC.

In this paper, we propose a novel quantum memory architecture designed to mitigate manufacturing defects in memory cells, thereby improving the yield of qRAM while lowering the degree of QEC to minimize the physical qubit overhead. Our proposed qRAM adopts a redundant repair scheme based on the Bucket Brigade architecture. By employing redundant qubits, we minimize the number of additional physical qubits for the QEC scheme. Moreover, by lowering the degree of QEC, we reduce the number of physical qubits per logical qubit and repair the resulting fabrication error using additional redundant qubits. This scheme considerably reduces the number of physical qubits constituting one qRAM. Here, we compare and analyze the yields for various error rates in fabricating physical qubits, various degrees of QEC applied to logical qubits, and the application of redundant repair technology. Through this analysis, we improve the yield to 95.92\% using only eight redundant qubits supported by an additional 1.01\% of all physical qubits. Our contributions are summarized below.
\begin{itemize}
    \item Firstly, we advocate circuit-level redundant repair for qRAMs, enabling fault recovery of defective logical qubits, and addressing fabrication error and QEC overhead.
    \item Secondly, we propose a qRAM architecture with a redundant repair scheme. The proposed architecture is based on the Bucket Brigade structure~\cite{arunachalam2015ot}, which is less vulnerable to noise than the Fanout structure. We introduce several key elements for the redundant repair scheme and support our proposed qRAM architecture with a quantum circuit model.
    \item Lastly, we analyze the qRAM yield in our method while varying the fabrication error rate and number of logical qubits for the qRAM. Moreover, we analyze the power of the effects of varying redundant qubit quantities on the qRAM yield. Furthermore, we evaluate the resource overhead by considering the necessary amount of physical qubits.
\end{itemize}

The remainder of this paper is organized as follows. In Sect. "Background and Motivation", we provide backgrounds of Bucket Brigade qRAM and quantum fabrication defects and motivate our research. In Sect. "Built-in Self Repair for qRAM", we explain our proposed architecture and present the redundant repair scheme. Sect. "Circuit-Model Implementation" describes our circuit model implementation. The experimental setup and analysis of performance evaluation are given in Sect. "Performance Evaluation". In Sect. "Related Works", we review prior related works, and in Sect. "Discussion", we provide a comparative discussion of our proposed method with related work addressing manufacturing defects in qRAM. Finally, we conclude our work in Sect. "Conclusion".

\section*{Background and Motivation}
\label{sec:background}
This section explains the Bucket Brigade qRAM architecture and quantum fabrication defects as well as the motivation of our study.

\subsection*{Bucket Brigade qRAM}
Most representative quantum algorithms assume that the data of the algorithm are loaded into the quantum processor through a quantum oracle called a black box~\cite{hann2021po}. For example, the Grover and the Harrow-Hassidim-Lloyd (HHL) algorithms can achieve a significant speedup through quantum oracle~\cite{grover1996af,duan_survey_2020}. In fact, qRAM plays a crucial role in the quantum oracles that most representative quantum algorithms assume~\cite{hann2021po}. qRAM can store quantum information in either the classical representation ($|0\rangle$ or $|1\rangle$) or the quantum representation (an arbitrary superposition of $|0\rangle$ and $|1\rangle$). qRAM also allows querying superposition of address form~\cite{di2020ft, arunachalam2015ot, hann2021ro}, as shown in Equation~\eqref{equ:qram}:
\begin{equation}
    \sum_j \alpha_j|j\rangle|0\rangle \stackrel{\mathrm{qRAM}}{\longrightarrow} \sum_j \alpha_j|j\rangle\left|m_j\right\rangle,
    \label{equ:qram}
\end{equation}
where $\sum_j \alpha_j|j\rangle$ is a superposition of query addresses and $\left|m_j\right\rangle$ represents the content of the $j$-$th$ memory location. qRAM can store the Grover algorithm~\cite{grover1996af}, HHL algorithm~\cite{duan_survey_2020}, and other quantum algorithms~\cite{adedoyin2018qa, montanaro2016qa, goss2022hf, leymann2020tb} as classical information while allowing superposed queries, thus offering considerable speedup over classical algorithms.

\begin{figure}[!t]
     \centering
     \begin{subfigure}[b]{0.47\textwidth}
         \centering
         \includegraphics[width=\textwidth]{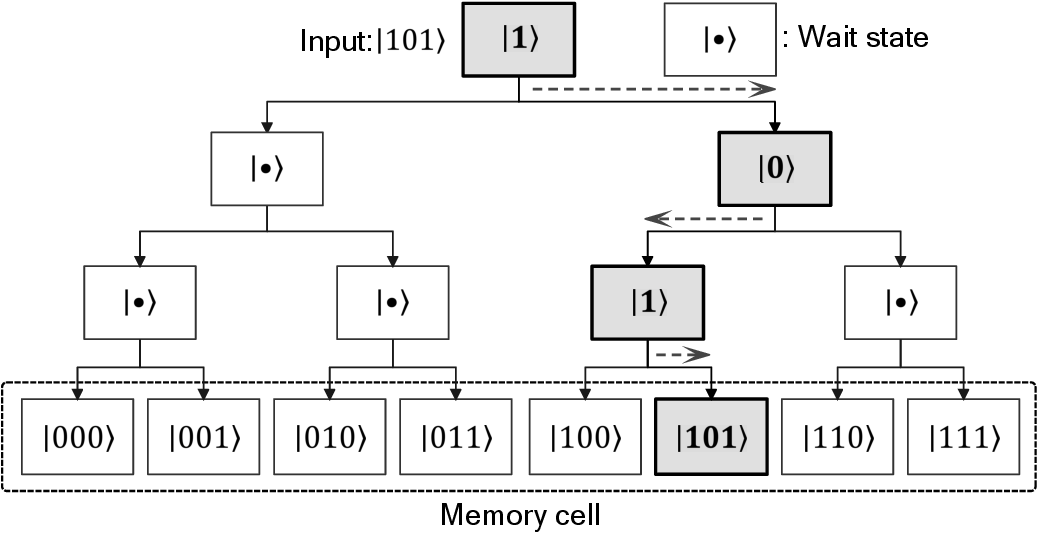}
         \caption{Bucket Brigade qRAM architecture. The routing nodes are qutrits with three possible states: $|0\rangle$, $|1\rangle$, and $|\bullet\rangle$ states. The $|\bullet\rangle$ state is the ``wait'' state, indicating that no activation is performed on that node. Input state is $|101\rangle$, the routing path is routed sequentially to the memory cell colored gray.}
         
         \label{fig:architectures}
     \end{subfigure}
     \hfill
     \vspace{0.3cm}
     \begin{subfigure}[b]{0.47\textwidth}
         \centering
         \includegraphics[width=\textwidth]{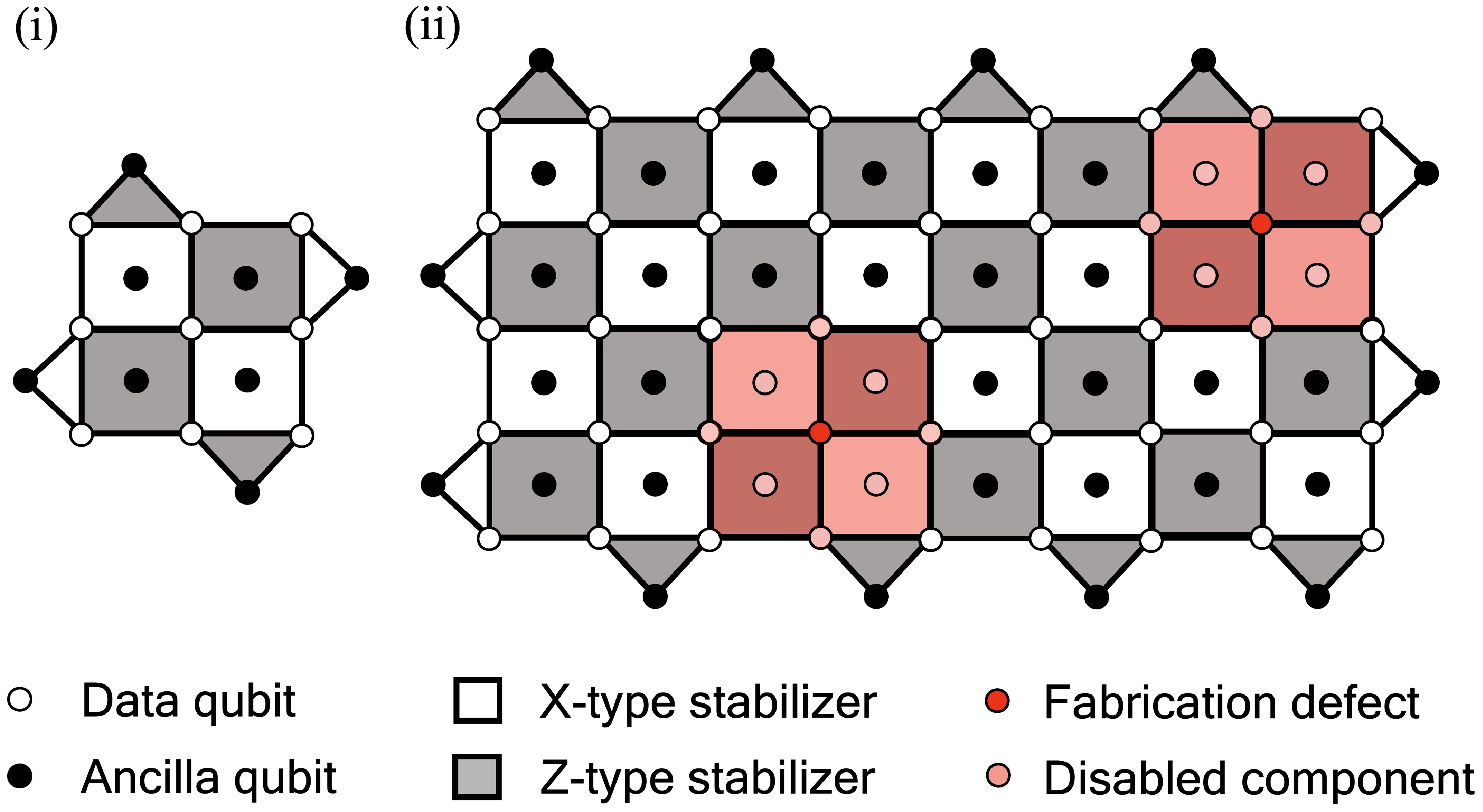}
         \caption{Fabrication defects in the surface code. Red circles are defective physical qubits introduced by fabrication errors. Both X- and Z-type stabilizers, data, and ancilla qubits connected to defective physical qubits (highlighted in pink) are disabled components.}
         \label{fig:qecc}
     \end{subfigure}
     \hfill
        \caption{Bucket Brigade qRAM architecture example (left) and occurrence of fabrication defects on surface code (right)}
        \label{fig:errorrate_yield}
\end{figure}

To describe the qRAM algorithm, we present a qRAM architecture model based on the Bucket Brigade architecture. \textcolor{REVISE_COLOR_2}{This architecture can support the superposition of addresses as input~\cite{phalak_quantum_2023}, but for better understanding, we show on the process of memory addressing using a specific input address in Figure~\ref{fig:architectures}.} This architecture can be routed in a three-level quantum-controlled `qutrit'~\cite{goss2022hf, arunachalam2015ot, hann2021ro, asaka2023tl} fashion. Each node of the binary tree is a `trit' with three possible internal states: $|0\rangle$, $|1\rangle$, or $|\bullet\rangle$. A trit in state $|0\rangle$ or $|1\rangle$ acts as a switch that routes the incoming signal to the `left' or `right', respectively. A trit in the $|\bullet\rangle$ state does not propagate the incoming signal. It continues until it reaches the $k$-$th$ level of the tree. The gray routing path in Figure~\ref{fig:architectures} is the path determined by the input state $|101\rangle$. Once the $n$ qubits of the input register $|101\rangle$ have passed through the tree, $n$ quantum switches are active, i.e., in the state $|0\rangle$ or $|1\rangle$; consequently, the output is $|{101}\rangle$ memory cell. Note that although this procedure requires the order of $2^n$ qutrits, only $n$ qutrits are active in any run of the protocol~\cite{di2020ft, arunachalam2015ot, hann2021ro, asaka2023tl}.

\subsection*{Quantum Fabrication Defects}
Quantum noise is the major problem limiting the computational advantage of current quantum computers. As quantum information is very fragile when encoded in larger quantum systems, the quantum states must be protected by QEC~\cite{egan2021ft, hann2021ro, suzuki2022qa, krinner2022rr}. Besides detecting and correcting errors, QEC provides an additional degree of freedom for controlling the complexity of logical qubit encoding. In practice, QEC observes the errors occurring on fault-tolerant circuits while attempting to control the logical qubits.

When constructing topological QEC codes, a lattice of physical qubits must be embedded on a manifold with a non-trivial topology such that the quantum information is encoded in the global degrees of freedom, as shown in Figure~\ref{fig:qecc}(i)~\cite{google2023sq, ni2023bt, krinner2022rr, jayashankar2022qe}. However, the industrial production line of large-scale topological devices introduces defective physical qubits because manufacturing processes are inherently imperfect. We refer to such faults as quantum fabrication defects~\cite{vovrosh2021sm, auger2017ft, tang2016rs}. In the example of Figure~\ref{fig:qecc}(ii), quantum fabrication defects are indicated as red-colored circles. If fabrication defects occur in the topology of the surface code lattice, the distances of the code and the quality of the encoded logical qubits are seriously reduced~\cite{auger2017ft, nagayama2017sc, piveteau2021em, suzuki2022qa}.

Fabrication defects also create punctures in the qubit array, some of which may be very large. We cannot assume control over the qubits within each puncture. Some protocols are designed to reliably collect the syndrome data near fabrication defects and perform the computation over the remaining intact qubits of the lattice with high probability, assuming a suitably low error rate of the intact qubits~\cite{auger2017ft, tang2016rs, nagayama2017sc}. However, if several defective physical qubits are sparsely distributed through the qubit array, this method is difficult to implement on a limited-size lattice. The lattice must then be expanded with additional qubits, which incurs a large resource overhead in terms of physical qubits.

\begin{figure}[!t]
     \centering
     \begin{subfigure}[b]{0.47\textwidth}
         \centering
         \includegraphics[width=\textwidth]{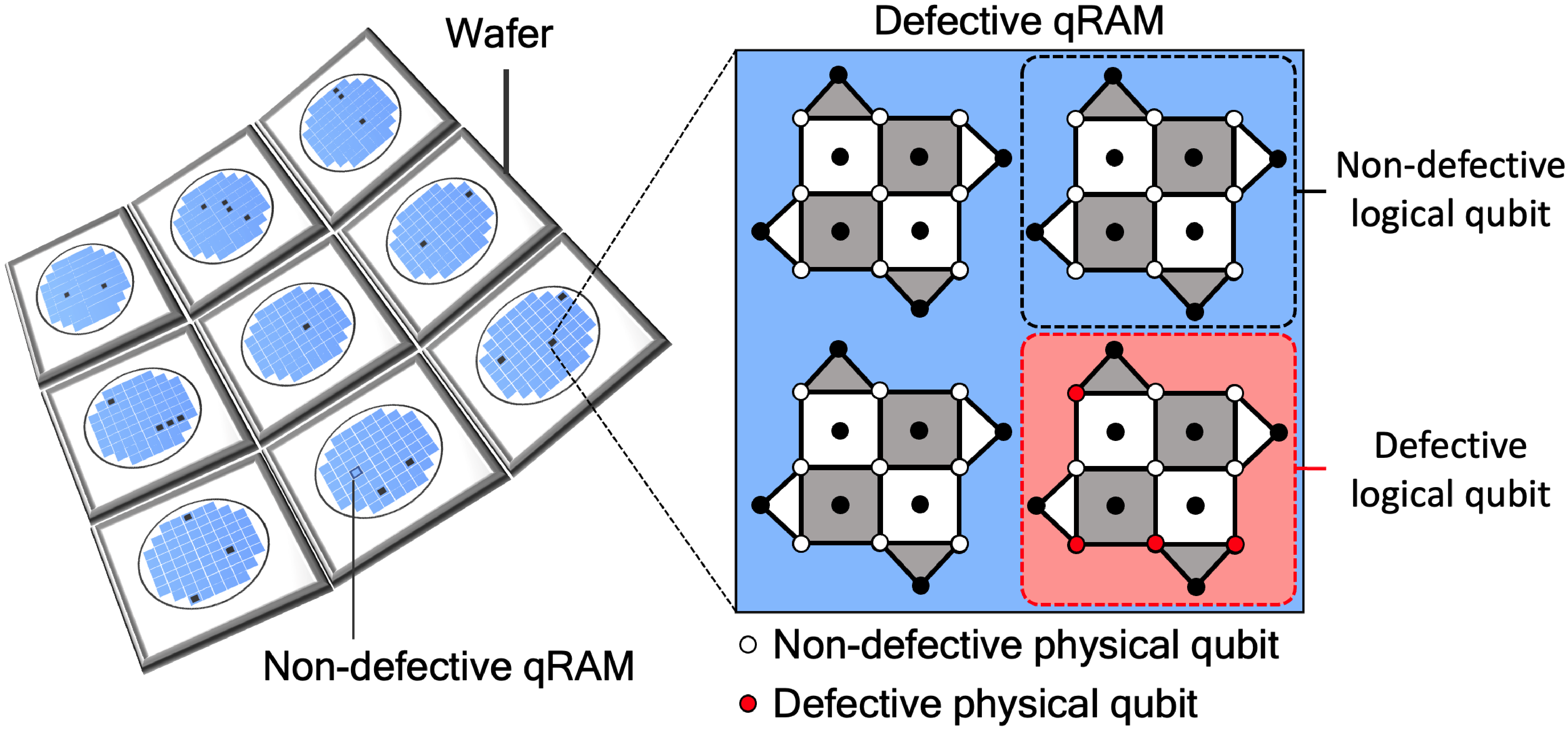}
         \caption{qRAMs on the wafer with defective logical and physical qubits}
         \label{fig:fabrication of qRAM}
     \end{subfigure}
     \hfill
     \vspace{0.3cm}
     \begin{subfigure}[b]{0.47\textwidth}
         \centering
         \includegraphics[width=\textwidth]{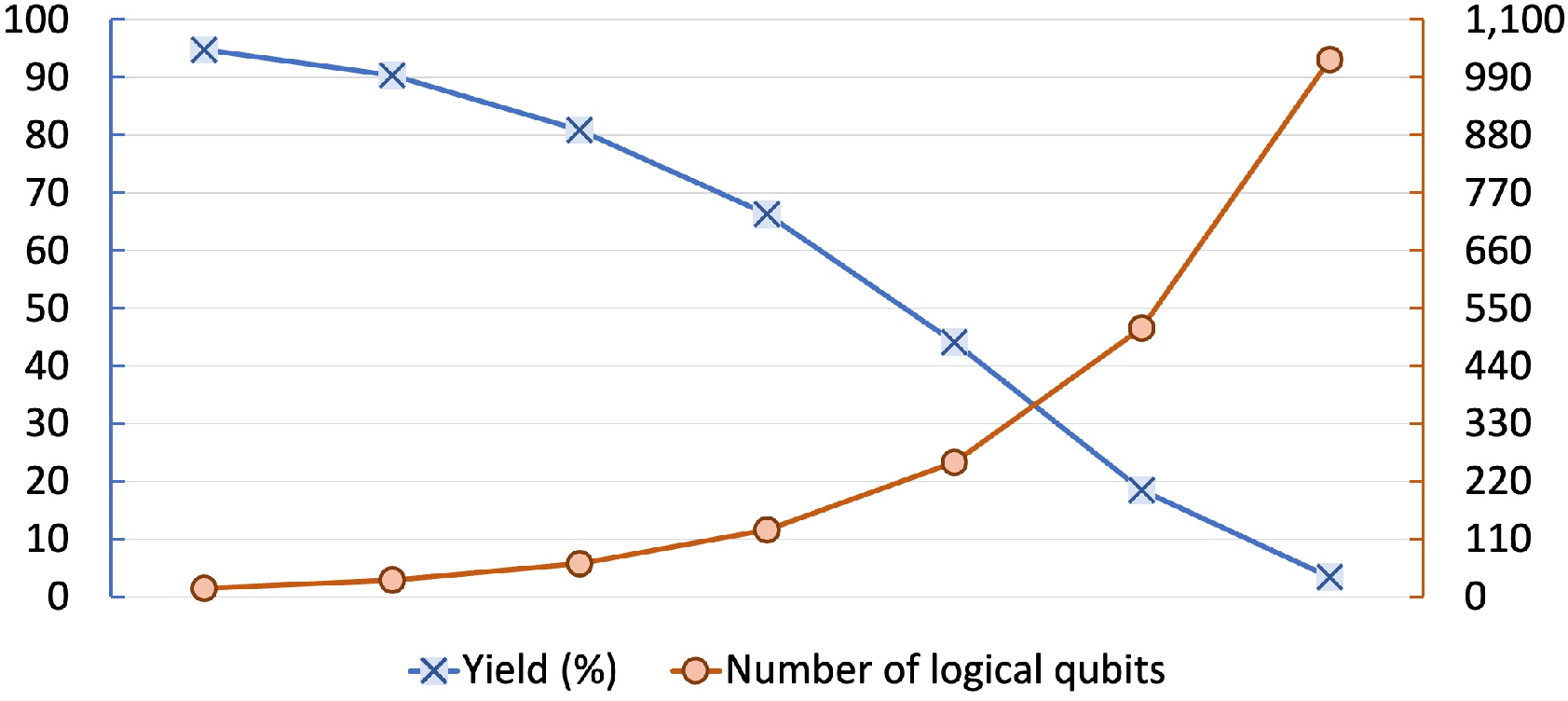}
         \caption{Relationship between the yield of qRAM and number of logical qubits}
         \label{fig:Yield with PQ}
     \end{subfigure}
     \hfill
        \caption{Fabrication of qRAM with wafers (left) and the relationship between the number of logical qubits qRAM yield (right)}
        \label{fig:fig2}
\end{figure}

\subsection*{Motivation}

qRAM is an essential element in practical quantum computation. In addition, qRAM must contain sufficiently many qubits for theoretical quantum calculations using practical-scale quantum algorithms that utilize large amounts of data. Recent improvements in silicon technology have increased the density and hence the number of qubits that can be manufactured. Practical-scale quantum algorithms and quantum gains are expected in the future, but we believe that a large number of qubits will be infected by fabrication errors. In fact, numerous fabrication-induced defects have been reported in superconducting quantum dots, which are promising candidates for building scalable quantum computers~\cite {bilmes2020rt, barends2013cj, verjauw2022pt}.

Such defects have fatal consequences in qRAM. If the memory cells constituting the qRAM include defective qubits, the qRAM cannot support random address accesses, and the qRAM yield is greatly reduced. Figure~\ref{fig:fabrication of qRAM} shows a qRAM fabricated with wafers, along with its logical and physical qubits. In the left upper panel of this figure, the non-defective qRAM cells (blue squares) are interspersed with defective qRAM cells (black squares). If one black square is magnified as shown in the upper right panel of Figure~\ref{fig:fabrication of qRAM}, we observe (for simplicity) four logical qubits encoded with multiple physical qubits. Among the four logical qubits is one defective qubit containing several defective physical qubits. Figure~\ref{fig:Yield with PQ} shows the simulated relationship between the yield and the number of logical qubits in the qRAM. The yield is calculated as Equation~\eqref{equ:yield}.
\begin{gather}
    \begin{split}
    {Yield}&=\left(1-\frac{{Number~of~defective~qRAMs}}{{Number~of~fabricated~qRAMs}}\right)\times 100
    \end{split}
    \label{equ:yield}
\end{gather}
Our simulation was performed on 1,000 qRAMs. Each logical qubit in each qRAM was encoded with 17 physical qubits subjected to a 0.5\% fabrication error rate. As shown in Figure~\ref{fig:Yield with PQ}, the yield decreases with the increasing number of logical qubits in the qRAM.

Theoretically, the low qRAM yield can be mitigated with the QEC scheme. There is a lot of research that employs high-degree QEC for encoding a logical qubit, which improves the reliability and fault tolerance of logical qubits~\cite{fowler_surface_2012,bonilla_ataides_xzzx_2021,ueno_qecool_2021}. However, high-degree QEC incurs enormous resource overhead because it requires an exponentially increasing number of physical qubits. Given the limited resources for making physical qubits at present, a high-degree QEC scheme is expected to reduce the productivity and yield of qRAM. The number of physical qubits in a single qRAM is determined by Equation~\eqref{equ:number_of_PQ}:
\begin{gather}
    \begin{split}
    N_{PQ}(qRAM) &= N_{LQ}(qRAM)\times N_{PQ}(Logical~Qubit),
    \end{split}
    \label{equ:number_of_PQ}
\end{gather}
where $N_{PQ}(qRAM)$ and $N_{LQ}(qRAM)$ represent the numbers of physical and logical qubits in the qRAM, respectively, and $N_{PQ}(Logical~Qubit)$ is the number of physical qubits required for implementing a single logical qubit. In other words, the number of physical qubits in the qRAM is the product of the number of logical qubits in the qRAM and the number of physical qubits per logical qubit.
Therefore, we must improve the mass-production yield of qRAM while reducing the resource overhead, i.e., the number of required physical qubits per qRAM. For this purpose, we propose a redundant repair method using additional spare qubits in the Bucket Brigade qRAM structure. This method reduces the number of physical qubits required per qRAM by lowering the degree of QEC. It also maximizes the quality of the qRAM yield by replacing defective qubits with redundant qubits. \textit{To the best of our knowledge}, our proposed redundant repair scheme in qRAM is the first approach to improve qRAM yield while minimizing resource overhead.

\section*{Built-in Self Repair for qRAM}
\label{sec:overall architecture}
In this section, we present our novel qRAM architecture based on the Bucket Brigade structure and our redundant repair scheme that resolves quantum fabrication defects.
\subsection*{Overall Architecture}

\begin{figure*}[!t]
    \centering{\includegraphics[width=17.5cm]{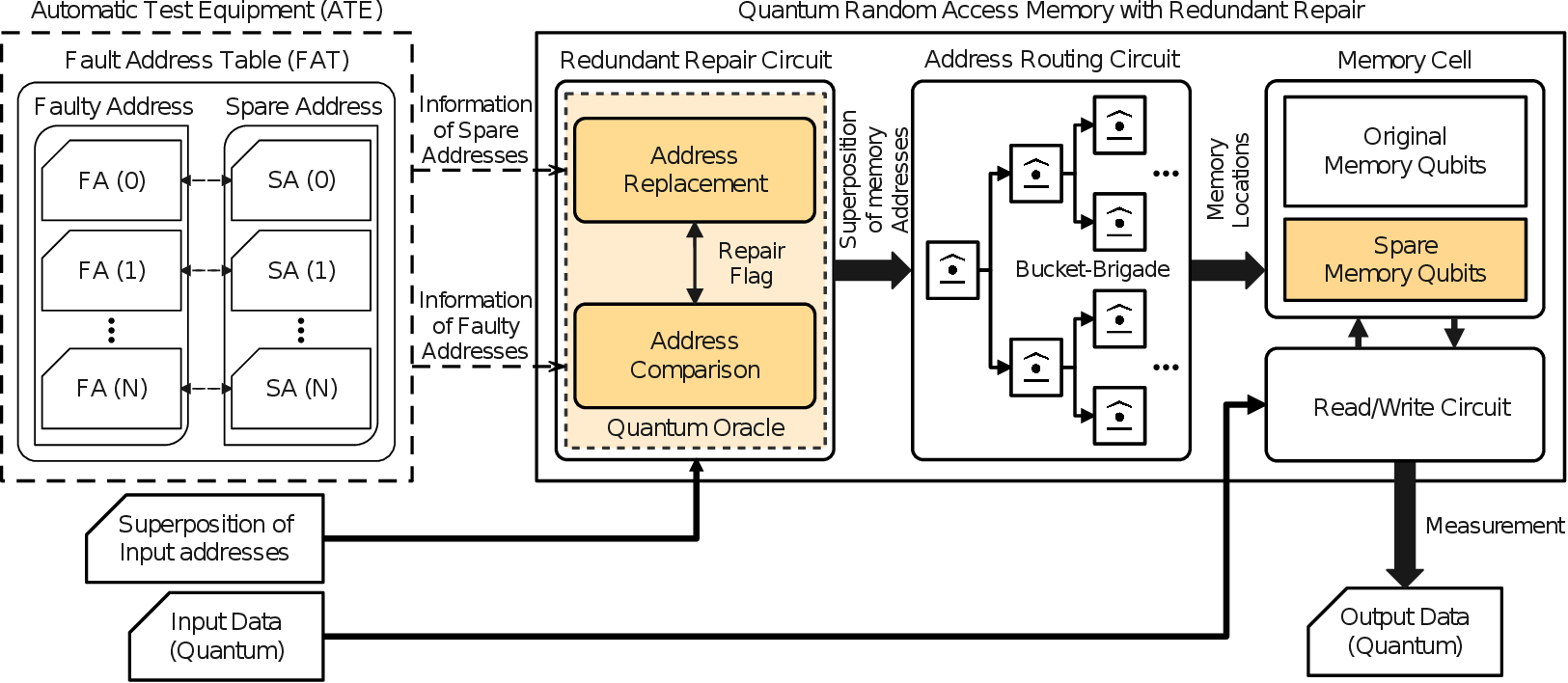}}
    \caption{~\textcolor{REVISE_COLOR_2}{Overall architecture of our proposed qRAM with redundant repair scheme and an external device ATE. All the data and states transferred to/from different parts are quantum. The ATE sends information of faulty addresses and spare addresses from the FAT to the quantum oracle of the redundant repair circuit. Based on two different types of address information, quantum oracle implements address comparison and address replacement parts, respectively. When the superposition of input addresses is given as input to qRAM, address comparison compares faulty addresses from FAT with each of the input addresses and sets the value of the repair flag. Checking the repair flag, the address replacement decides whether to replace or not replace the input address with the spare address. After replacing all faulty addresses, the redundant repair circuit passes the superposition of memory addresses to the address routing circuit for routing. Memory locations are then given to the memory cells, which communicate with read/write circuit data to read the memory cell data (Quantum) and write the input data (Quantum) to the memory cell.}}
    \label{fig:workflow}
\end{figure*}

Figure~\ref{fig:workflow} shows the overall architecture of our proposed qRAM. The core of our architecture lies in the redundancy repair scheme, which functionally bypasses memory cells with fabrication defects by utilizing spare qubits. This repair mechanism assumes that defect information is provided by automatic test equipment (ATE) and a fault address table (FAT). The ATE is a classical, standard industrial equipment used to test the functionality and performance of chips after the semiconductor manufacturing process. In the classical RAM manufacturing process, the ATE tests the memory array to identify the addresses of non-operational (defective) cells. Based on these test results, the ATE generates an information table called the fault address table, which maps the addresses of defective cells (faulty address, $FA$) to the addresses of spare cells (spare address, $SA$). This table is typically stored in fuses or non-volatile memory so that a built-in self-repair mechanism can reference it~\cite{fang2003em, sridhar2012bi,1270863}, and it is used to bypass fault addresses during memory operation.

The focus of our work is not on the development of ATE technology for detecting defects in qRAM. We assume that the standard approach from the classical RAM industry can be applied to future quantum device manufacturing processes. In fact, studies such as \cite{bilmes2020rt, PhysRevA.109.022440} are actively being conducted to detect and identify fabrication defects in quantum hardware. Based on these research trends, we assume that test equipment, such as ATE, will successfully perform defect identification and FAT generation to provide the defect information. Under this premise, our key contribution is to propose a resource-efficient BISR architecture mechanism that utilizes the pre-identified defect information from the ATE to bypass fabrication defects that occurred during the manufacturing process.

Our proposed qRAM consists of four major components: the redundant repair circuit, address routing circuit, read/write circuit, and memory cells.

First, the redundant repair circuit of qRAM receives the information of faulty and spare addresses from the ATE. Based on these addresses, the redundant repair circuit supports a quantum oracle that consists of address replacement and address comparison modules. When the superposition of input addresses is given to the quantum oracle, the address comparison module determines whether the input address has an equality match with a known faulty address stored in the FAT. The result of this comparison operation sets the state of the repair flag qubit. If the input address matches the faulty address, the flag's state is set to $|1\rangle$. The state of this repair flag is useful because it acts as a switch that directly determines whether the address replacement module is activated. If the flag is in the $|1\rangle$ state (defect match), the address replacement module is activated and performs the function of replacing the faulty address with the spare address. Conversely, if the flag remains in its initial $|0\rangle$ state (no match), the address replacement module remains bypassed, allowing the original input address to pass through unchanged. This process is repeated until all superposition of input addresses is checked. After all iterations, the superposition of memory addresses comes out as the output of the redundant repair circuit.

Second, the address routing circuit activates the routing nodes with the superposition of the memory address. The structure of the address routing circuit is based on Bucket Brigade qRAM, with the routing nodes configured as a binary tree. From the initial state $|\bullet\rangle$ of each routing node, specific routing nodes are activated by setting $|0\rangle$ or $|1\rangle$ based on the superposition of memory addresses. The redundant repair circuit provides the locations of original and spare memory qubits. Since memory addresses are in the superposed state, they address all memory cells with the same probability except defected memory cells.

Third, the read/write circuit reads the data in the memory cells or writes data to the memory cells. The read/write circuit can access the memory cells with memory location data. When reading from memory, the read/write circuit finds the memory locations of the memory cells and accesses them for measurement. After the measurement, the read/write circuit generates the output data. When writing to memory, the read/write circuit accesses the memory cells with the memory locations and writes data to those cells. At this time, the read/write circuit must be provided with input data. 

Finally, in memory cells, there are original and spare memory qubits. By adopting spare memory qubits, we can achieve redundant repair for memory cells if faults occur on original memory cells. Based on the data of memory locations from the address routing circuit and data from the read/write circuit, qRAM reads or writes both types of qubits avoiding addressing defective ones.

\subsection*{Redundant Repair Scheme}

We clearly describe the operational principle of the redundant repair scheme for qRAM. The core of this scheme is the one-to-one replacement of an input address with a pre-designated, dedicated spare address ($SA$) if the input address is a faulty address ($FA$).

First, we define the main components used in the process. The fault address table is defined as a set of $n$ ordered pairs:

\begin{equation}\label{equ:FAT}
   \begin{aligned}
   FAT = \{ (FA_i,~SA_i)~|~i = 1, 2, ... , n \}
   \end{aligned}
\end{equation}

Here, $FA_i$ represents the $i$-th faulty address, and $SA_i$ represents the dedicated spare address pre-allocated to replace $FA_i$ when it is detected.
The input is given as a superposition of input addresses ($SoI$).
The redundant repair scheme applies an address transformation function called $Repair$ to each individual address state within this superposition. For this function, we define $A_{in}$ as an arbitrary single input address (i.e., one state from $SoI$), and $A_{out}$ as the final memory address that the $Repair$ function maps $A_{in}$ to.

The transformation logic is determined by checking the FAT. If $A_{in}$ matches a faulty address $FA_i$ registered in the FAT, the redundant repair process immediately replaces $A_{in}$ with the corresponding $SA_i$:



    \begin{equation}\label{equ:1st_case}
        \begin{aligned}
    A_{out} = SA_i, \text{ if } A_{in} = FA_i \text{ and } (FA_i, SA_i) \in FAT
        \end{aligned}
    \end{equation}

This one-to-one mapping ensures that the replacement operation is not applied recursively. When $A_{in}$ matches $FA_i$, the output is deterministically set to $SA_i$. The process does not subsequently check if $SA_i$ is equivalent to another fault address (e.g., $FA_j$), thereby preventing any chained replacements.
Otherwise, if $A_{in}$ does not match any faulty address registered in the FAT, it is treated as a non-faulty address and passes through unchanged.

    \begin{equation}\label{equ:2nd_case}
        \begin{aligned}    
    A_{out} = A_{in}, \text{ if } A_{in} \notin \{ FA_1, \dots, FA_n \}
        \end{aligned}
    \end{equation}

Integrating these two cases, the $Repair$ function can be defined as equation~\eqref{equ:final_equ}.

    \begin{equation}\label{equ:final_equ}
        \begin{aligned}
    A_{out} = Repair(A_{in}) = \begin{cases} SA_i & \text{if } \exists i \text{ such that } (A_{in} = FA_i) \text{ and } (FA_i, SA_i) \in FAT \\ A_{in} & \text{otherwise} \end{cases}
            \end{aligned}
    \end{equation}

Finally, we show how this $Repair$ function operates on the entire superposition, extending the ideal qRAM operation from equation~\eqref{equ:FAT}. Based on the static information from FAT (equation~\eqref{equ:FAT}), we can logically partition the total input state. We define $qRAM_{RR}$ as the operation of our proposed $Repair$ logic shown in equation~\eqref{equ:final_equ}. This entire redundant repair process, showing the $qRAM_{RR}$ operation on the partitioned input state, is expressed as equation~\eqref{equ:qRAMrr}:

\begin{equation}\label{equ:qRAMrr}
    \begin{aligned}
        \left( \sum_{j \in J_{Good}}\alpha_{j}|j\rangle + \sum_{k \in J_{Defect}}\alpha_{k}|k\rangle \right) |0\rangle \xrightarrow{qRAM_{RR}} \sum_{j\in J_{Good}}\alpha_{j}|j\rangle|m_{j}\rangle+\sum_{k\in J_{Defect}}\alpha_{k}|SA_{k}\rangle|m_{SA_{k}}\rangle
    \end{aligned}
\end{equation}

Here, the $qRAM_{RR}$ operation acts in parallel on both partitions due to quantum linearity. It applies the logic of equation~\eqref{equ:2nd_case} to the $J_{Good}$ partition and the logic of equation~\eqref{equ:1st_case} to the $J_{Defect}$ partition. This yields the final output state where the address register (now representing the $SoM$) is entangled with the data register, and all faulty addresses have been successfully re-routed to their corresponding spare addresses.

Algorithm~\ref{alg:rr_qram} shows the procedural logic that implements the non-recursive transformation shown in the final state of equation~\eqref{equ:qRAMrr}. This classical pseudocode abstraction treats $SoI$ as a list of its basis states (addresses) for clarity. The algorithm's core logic iterates over each input address ($SoI_j$). For each individual $SoI_j$, it performs a search through the entire $FAT$. If a match is found with a faulty address ($FA_i$), the algorithm immediately adds the corresponding spare address ($SA_i$) to the output $SoM$ and stops searching the $FAT$ for that specific $SoI_j$. This immediate termination (implemented via the $found\_match$ flag and $break$ command) is a crucial feature. It ensures that the replacement operation is non-recursive. This logic intrinsically prevents the problematic scenario where a spare address (e.g., $SA_i$) that just replaced a faulty one could itself be treated as a new faulty address and be replaced again. If, after checking the entire $FAT$, no match is found for $SoI_j$, the original address is considered non-faulty and is added to $SoM$, implementing the logic of equation~\eqref{equ:1st_case}.

\begin{algorithm}[!tbh]
\caption{Redundant repair on qRAM}
\label{alg:rr_qram}
\begin{algorithmic}[1]
    \State \textbf{INPUT:} $FAT$: Fault address table, $SoI$: Superposition of Input addresses
    \State \textbf{OUTPUT:} $SoM$: Superposition of Memory addresses
    \newline
    
\Procedure{RedundantRepair}{$FAT$, $SoI$}

    \State $SoM \gets \emptyset$
    \For{$j \gets 1$ to \Call{size}{$SoI$}}
        \State $found\_match \gets \text{false}$
        \For{$i \gets 1$ to $n$}
            \If{$SoI_j = FA_i$}
                \State $SoM \gets SoM \cup \{SA_i\}$
                \State $found\_match \gets \text{true}$
                \State \textbf{break}
            \EndIf
        \EndFor
        \If{$found\_match = \text{false}$}
            \State $SoM \gets SoM \cup \{SoI_j\}$
        \EndIf
    \EndFor
    \State \Return $SoM$
\EndProcedure
\end{algorithmic}
\end{algorithm}




\begin{figure}[!ht]

    \centering{\includegraphics[width=0.455\textwidth]{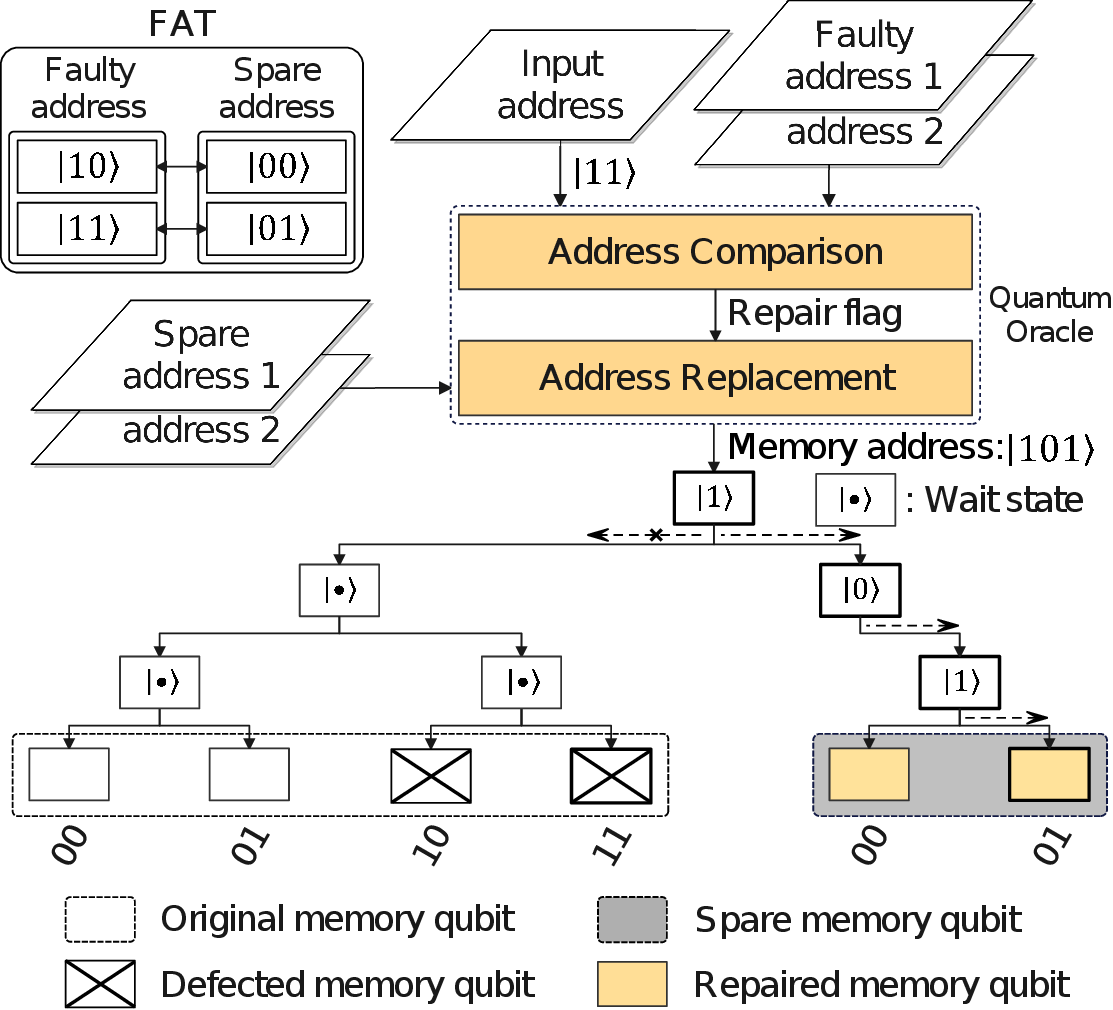}}
    \caption{\textcolor{REVISE_COLOR_2}{Example of our proposed redundant repair algorithm. When the superposition of addresses including faulty addresses is given as input the quantum oracle does address comparison and address replacement based on FAT. The output of the quantum oracle is the superposition of addresses as well. For faulty addresses, the repair flag qubit is activated to route spare memory qubits. original memory qubits will be routed.}}
    \label{fig:algorithm example}
\end{figure}

\begin{figure*}[!t]

    \centering{\includegraphics[width=17.5cm]{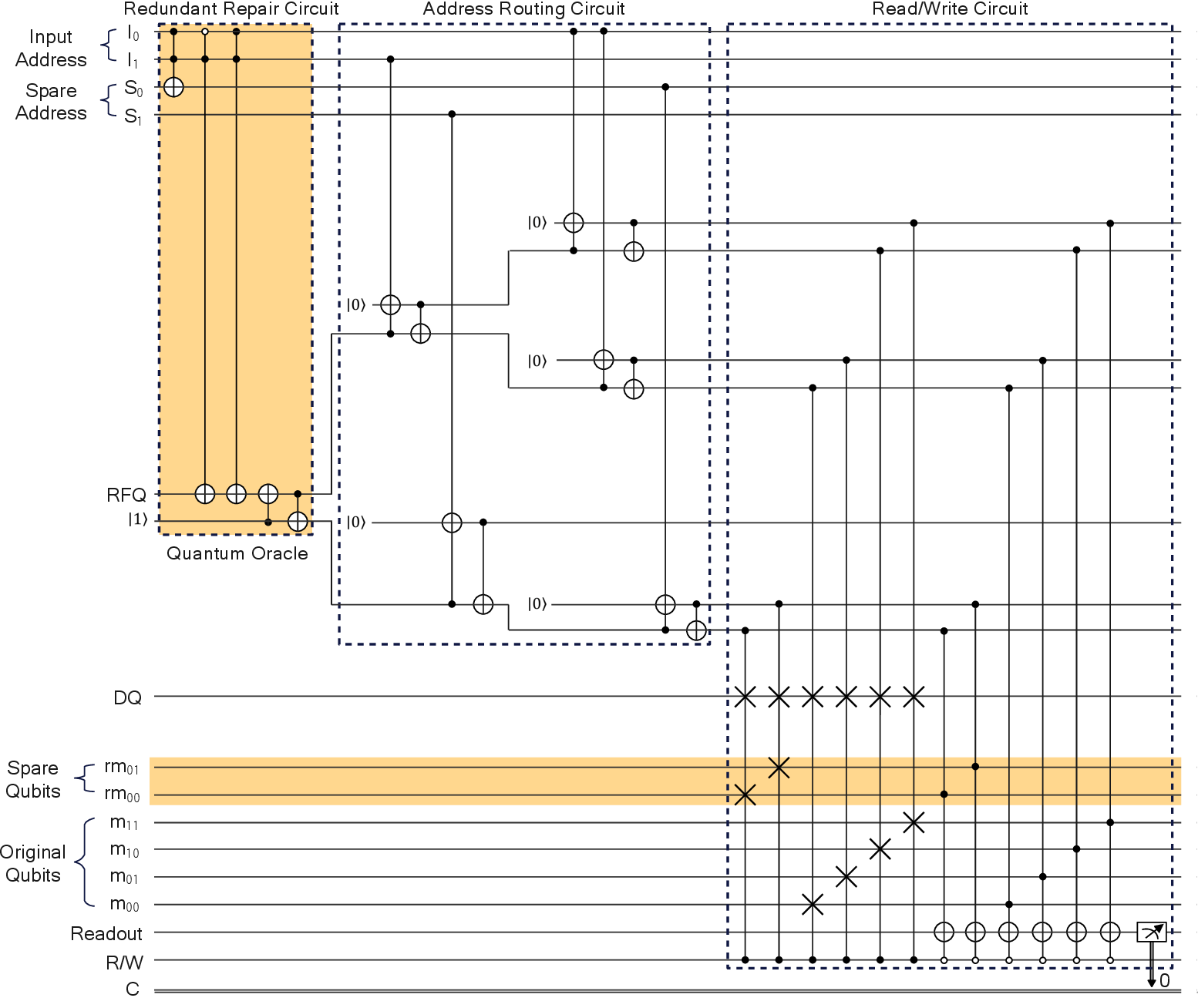}}

    \caption{Quantum circuit example of the proposed qRAM consisting of redundancy recovery circuit, address routing circuit, read/write circuit, and memory cell. This circuit is an example of an input address represented by two logical qubits. Parts highlighted with a yellow background are essential to support redundant repair schemes.}
    \label{fig:proposed architecture circuit model}
\end{figure*}

Figure~\ref{fig:algorithm example} is an example of the redundant repair algorithm with the Bucket Brigade qRAM. The FAT includes faulty addresses with their corresponding spare addresses. Since faulty addresses and spare addresses of FAT depend on the location of defects in memory cells, we use quantum oracle. We designed the quantum oracle as the Equation~\eqref{equ:oracle}.
~\textcolor{REVISE_COLOR_2}{The superposition of input addresses is entered into our quantum oracle. Each of the input addresses will be compared with faulty addresses in the FAT by the address comparison module. If two addresses match, the repair flag address is activated, which in turn causes the replacement module to replace the faulty address with the corresponding spare address based on the FAT. This process is repeated in the quantum oracle until all faulty addresses are replaced. After all iterations, the superposition of memory addresses will be given to the Bucket Brigade structure. In the figure, we simply represent $SoM$ as $|+\rangle$. There are no longer faulty addresses in $SoM$ so it routes all memory cells including spare ones except $|10\rangle$ and $|11\rangle$ states.} 

\begin{equation}\label{equ:oracle}
    U_{FA}|IA\rangle= 
\begin{cases}
    |RFQ\rangle \otimes |SA\rangle, & \text{if } IA = FA\\
    |RFQ\rangle \otimes |IA\rangle, & \text{otherwise }
\end{cases}
\end{equation}


\section*{Circuit-Model Implementation}
\label{sec:Circuit model implementation}

In this section, we describe the quantum circuit model of our proposed qRAM architecture. As the redundant repair scheme in the proposed method uses redundant qubits, some components of our scheme mirror those of Bucket Brigade in~\cite{arunachalam2015ot}. Figure~\ref{fig:proposed architecture circuit model} shows the entire quantum circuit model of our proposed qRAM. We assume there are four original memory qubits ($m_{00}$, $m_{01}$, $m_{10}$, and $m_{11}$) and two spare memory qubits ($rm_{00}, rm_{01}$) for the memory cells. We also employ two qubits each for the input address ($I_0, I_1$) and spare address ($S_0, S_1$). Since qRAM is designed to support when the input is given as a superposition state, we describe our proposed method assuming that the superposition of input addresses is given as input.

To illustrate the operation and defect management of this circuit model, we explain with a specific example. Consistent with the focus of this paper, we assume all circuit components (qubits, gates) operate ideally. The error to be managed is the pre-existing fabrication defect in a memory cell, not dynamic operational faults. The quantum oracle implements the repair logic by referencing the mapping defined in the FAT. This example uses the specific mapping $FA = |10\rangle$ (faulty address) to $SA = |00\rangle$ (spare address).

The proposed quantum circuit consists of four parts: a redundant repair circuit, an address routing circuit, a read/write circuit, and memory cells.
The redundant repair circuit includes multiple positive and negative control gates to support the quantum oracle implementation. This circuit compares each address in the superposition state with the fault addresses implemented in our quantum oracle. For example, when the non-faulty state $|01\rangle$ enters the oracle, the address comparison logic finds no match with $FA=|01\rangle$. Consequently, the repair flag qubit ($RFQ$) remains in its initial $|0\rangle$ state, and the address $|01\rangle$ passes through the address replacement module unchanged. Conversely, when the faulty state $|10\rangle$ enters, the address comparison logic detects a match. This match activates the control gates, flipping the $RFQ$ to $|1\rangle$. The address replacement logic is controlled by $RFQ$ and then replaces the faulty address $|10\rangle$ with the pre-defined spare address $|00\rangle$. All qubits for spare address are initialized with $|0\rangle$.

The address routing circuit is configured as a binary tree to mimic the Bucket Brigade structure. It consists of an upper binary tree for original memory qubits and a lower binary tree for spare memory qubits. The $RFQ$ acts as the primary root node (or switch) to select which tree is activated. Following our example, for the non-faulty state $|01\rangle$, the $RFQ$ remains $|0\rangle$. This state activates the upper binary tree. The address $|01\rangle$ is then passed to this tree, activating the routing nodes (entangled via $Toffoli$ and $CNOT$ gates) that lead to the original qubit $m_{01}$. In the faulty state $|10\rangle$ (the error management process), the $RFQ$ is flipped to $|1\rangle$. This $|1\rangle$ state deactivates the upper binary tree (preventing access to the defective $m_{10}$) and activates the lower binary tree. The newly replaced address $|00\rangle$ is then routed through this lower tree, activating the path to the spare qubit $rm_{00}$. Note that figure~\ref{fig:proposed architecture circuit model} is a specific example illustrating the case for only two spare qubits ($rm_{00}$ and $rm_{01}$). Therefore, the lower binary tree is constructed only to address these two physically implemented spare qubits, and its size is not $2^n-1$ but rather depends on the actual number of spare qubits being implemented (in this case, two).

The read/write circuit reads (writes) data from (to) memory. For this purpose, we entangle an $R/W$ qubit with routing nodes, a $DQ$ as the input data, and memory qubits through multi-controlled gates. Continuing our example, the circuit accesses the memory cell selected by the routing process. If the $R/W$ is set to read (e.g., $|0\rangle$), the circuit will access $m_{01}$ for the $|01\rangle$ state, or it will access $rm_{00}$ for the $|10\rangle$ state. The defect in $m_{10}$ is thus successfully managed by bypassing it. When the $R/W$ state is $|0\rangle$, the $Readout$ qubit is entangled with the routed memory cells and measured. Conversely, for memory writing (e.g., $R/W$ = $|1\rangle$), the $DQ$ (input data) is written to the routed memory cells ($m_{01}$ or $rm_{00}$). While this trace used classical basis states for clarity, the quantum circuit performs the entire logic of comparison, replacement, and re-routing in parallel for all states in the input superposition due to quantum linearity.

\section*{Performance Evaluation}
\label{sec:performance evaluation}
This section evaluates the yield and resource overhead of our proposed method. We first explain the experimental setup and evaluation metrics of our simulation and then analyze the performance in terms of the evaluation metrics. We also present resource overhead data and a simulated result table for a deeper level of understanding.

\subsection*{Experimental Setup}

We simulated the qRAM with a number of logical qubits from a minimum of 16 to a maximum of 1,024 while doubling its number of logical qubits. In addition, each logical qubit is constructed as a surface code lattice using $2d^2-1$ number of physical qubits. Here, $d$ is the code distance of the surface code, and $ \lfloor \frac{d-1}{2} \rfloor $ is the amount of error that can be corrected. With the code distance, we define a logical qubit as defective if it has more errors than it can correct. We also define a qRAM as defective if at least one logical qubit in the qRAM is defective. To vary the degree of QEC, we set $d$ to 3, 5, 7, and 9.
The $d$-dependent degree of QEC is denoted as 'QEC d'. 

For modeling the fabrication error occurrence to each physical qubit, we randomly injected errors with a binomial distribution at the fabrication error rate. The simulated fabrication error rate ranged from 0.5\% to 1\% in 0.1\% increments. As the fabrication error rate increases, the occurrence probability of defective logical qubits increases as well. We also evaluate yield improvement by varying numbers of redundant qubits to 1, 2, 4, and 8.

The simulation results were evaluated regarding two metrics: yield and resource overhead of the qRAM. The yield was computed using Equation~\eqref{equ:yield}. To determine the resource overhead, we must consider the total number of physical qubits required for constructing the proposed qRAM. To check the total number of physical qubits, we divided the memory cell from the peripheral parts (namely, the addressing qubits and qubits of the routing nodes).

To ensure the reliability and validity of our results, we ran the experiment 10 times with a total of 1,000 qRAMs. The yields are reported as the averages of the 10 experimental yields. The simulation results were obtained using an in-house simulation program for this study.

\subsection*{Performance Analysis}
\subsubsection*{Yield Improvement}
\begin{figure*}[!ht]
    \centering{\includegraphics[width=17.5cm]{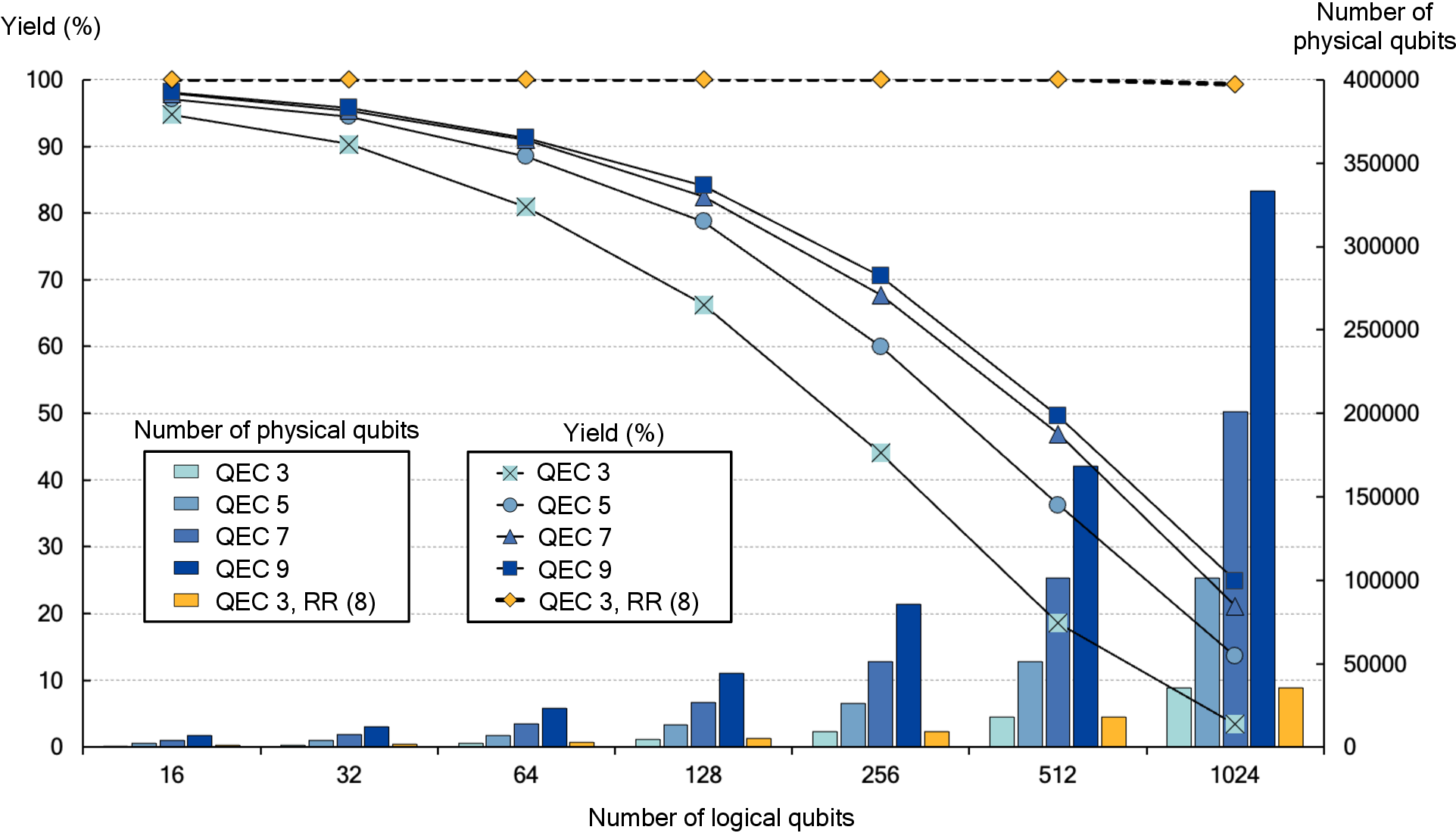}}
    \justifying
    \caption{Simulated yield (left y-axis) and number of physical qubits in the qRAM construction (right y-axis) versus number of logical qubits in qRAMs with different degrees of quantum error correction (QEC). The fabrication error rate was fixed at 0.5\%. $RR(8)$ denotes using eight spare qubits for redundant repair.}
    \label{fig:simulation result}
\end{figure*}

Figure~\ref{fig:simulation result} plots the qRAM yields and number of physical qubits required for qRAMs with and without redundant repair qubits for different degrees of QEC. 
The \textit{QEC 3, RR(8)} scheme in Figure~\ref{fig:simulation result} describes the QEC 3 scheme for logical qubits using eight additional redundant qubits as the spare memory qubits. A 0.5\% fabrication error rate was applied to all physical qubits in each logical qubit. As shown in the figure, the yield of a qRAM with a given number of logical qubits increased with an increasing degree of QEC. In the case of 256 logical qubits, the yields at QECs of 3, 5, 7, and 9 were 44.08\%, 60.0\%, 67.73\%, and 70.6\%, respectively. Moreover, the yield of the qRAM reached 100\% when employing eight redundant qubits with QEC 3. Meanwhile, the number of additional physical qubits required for the eight redundant qubits was 4.51\% of the total number of physical qubits utilized in QEC 3. By using a negligible additional number of physical qubits, the yield of qRAM was greatly increased. This outcome holds great significance since the utilization of a low degree of QEC with redundant qubits can substantially reduce the number of additional physical qubits rather than using a higher degree of QEC without redundant qubits to increase the yield of qRAM.


%

Through the results of Figure~\ref{fig:simulation result}, we demonstrated the average yield improvement according to the presence or absence of redundant repair. To obtain average yield improvement, we first calculated the average yields of QEC 3, QEC 5, QEC 7, and QEC 9 and differentiated them from the yield with the redundant repair. We defined this variation value as the average improvement of the yield for different numbers of logical qubits. Through this calculation, we illustrated the average yield improvement by 3.05\%, 6.01\%, 12.08\%, 22.09\%, 39.39\%, 62.14\%, and 83.59\% for 16, 32, 64, 128, 256, 512, and 1,024 logical qubits, respectively.

\begin{figure*}[!t]
     \centering
     \begin{subfigure}[b]{0.3\textwidth}
         \centering
         \captionsetup{justification=centering}
         \includegraphics[width=\textwidth]{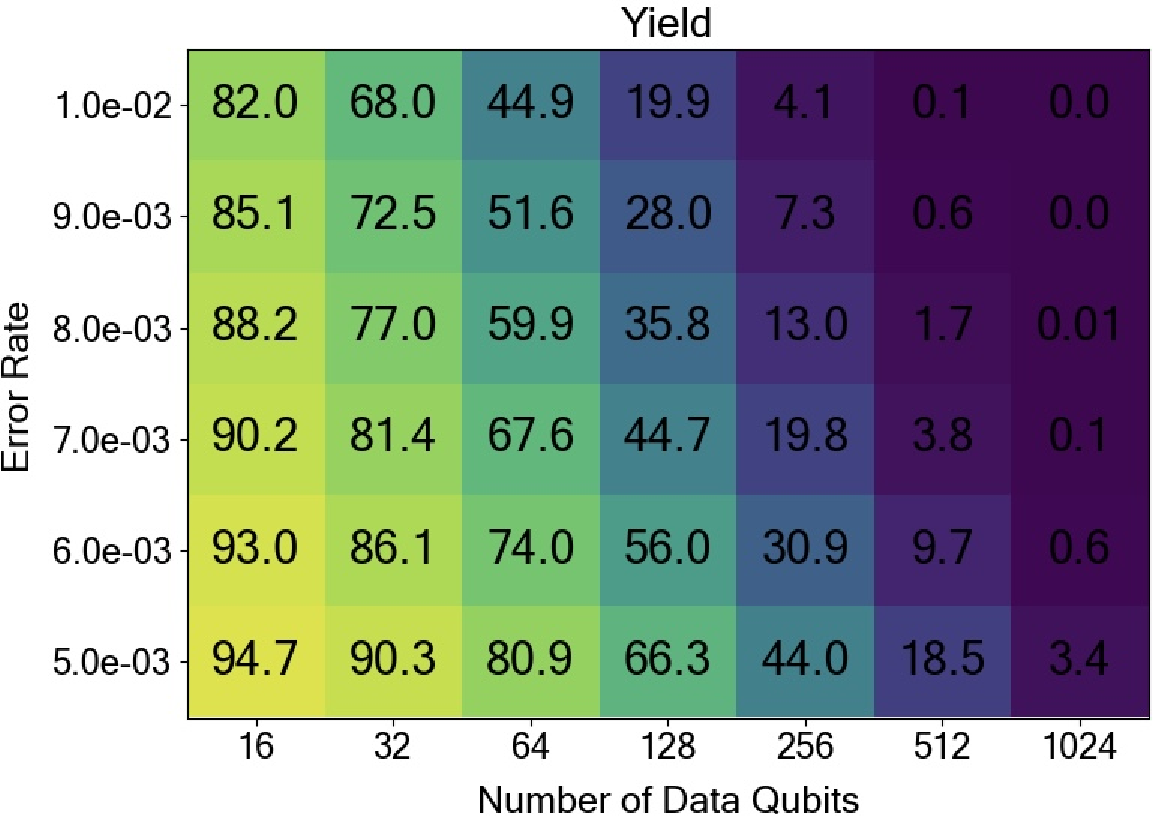}
         \caption{No redundant qubits}
         \label{fig:y equals x}
     \end{subfigure}
     \begin{subfigure}[b]{0.3\textwidth}
         \centering
         \captionsetup{justification=centering}
         \includegraphics[width=\textwidth]{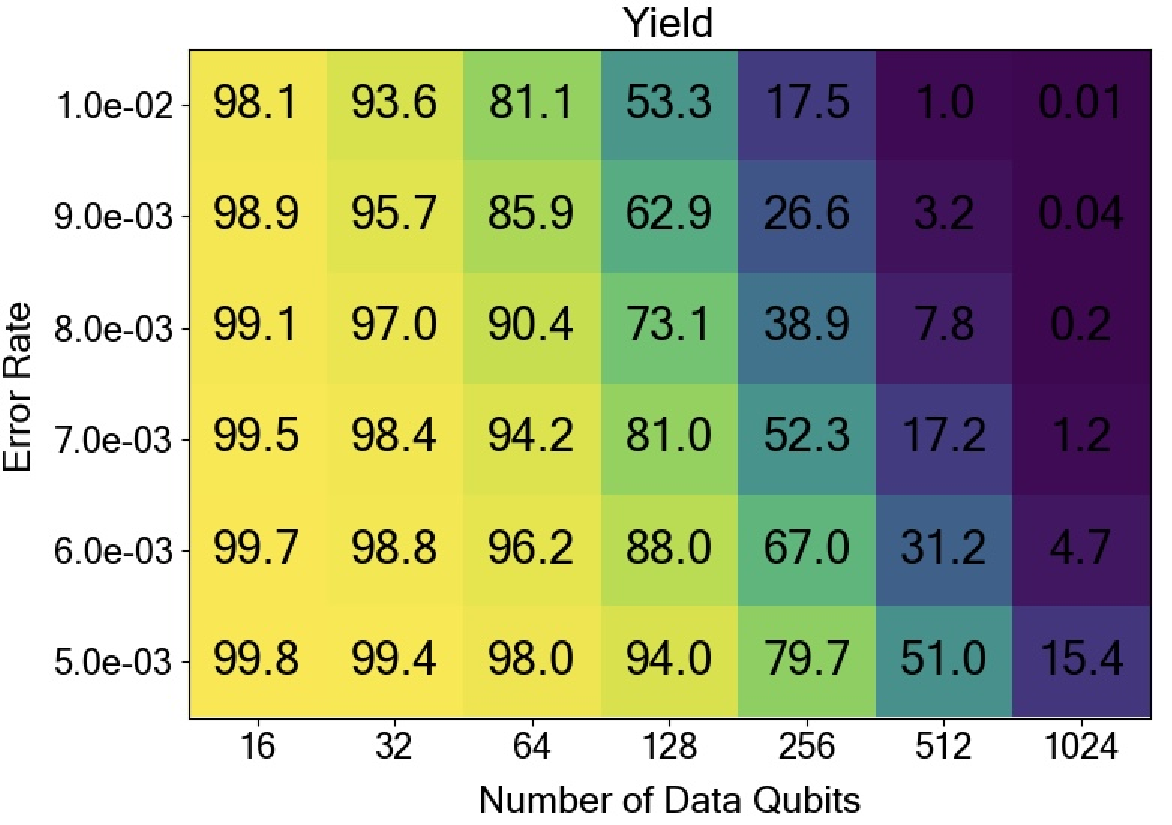}
         \caption{One redundant qubit}
         \label{fig:three sin x}
     \end{subfigure}
     \begin{subfigure}[b]{0.3\textwidth}
         \centering
         \captionsetup{justification=centering}
         \includegraphics[width=\textwidth]{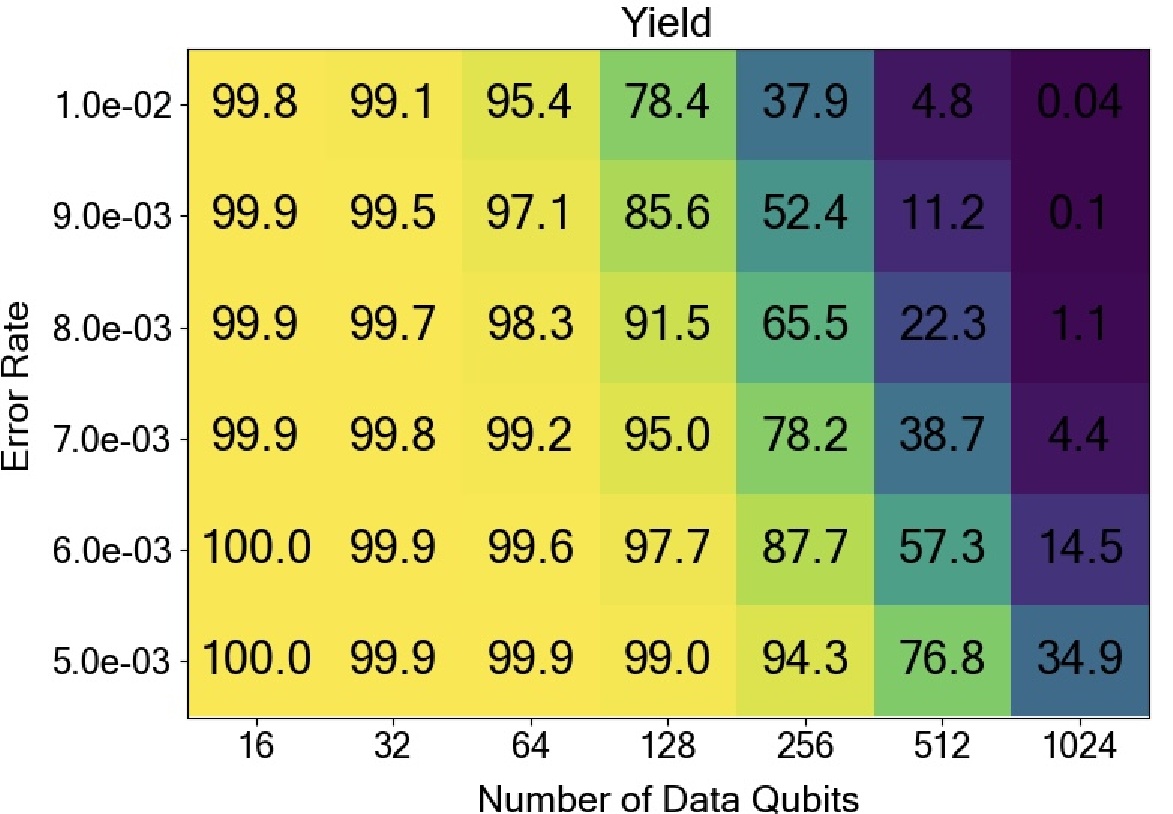}
         \caption{Two redundant qubits}
         \label{fig:five_Yield_rr2_v4}
     \end{subfigure}
     \begin{subfigure}[b]{0.3\textwidth}
         \centering
         \captionsetup{justification=centering}
         \includegraphics[width=\textwidth]{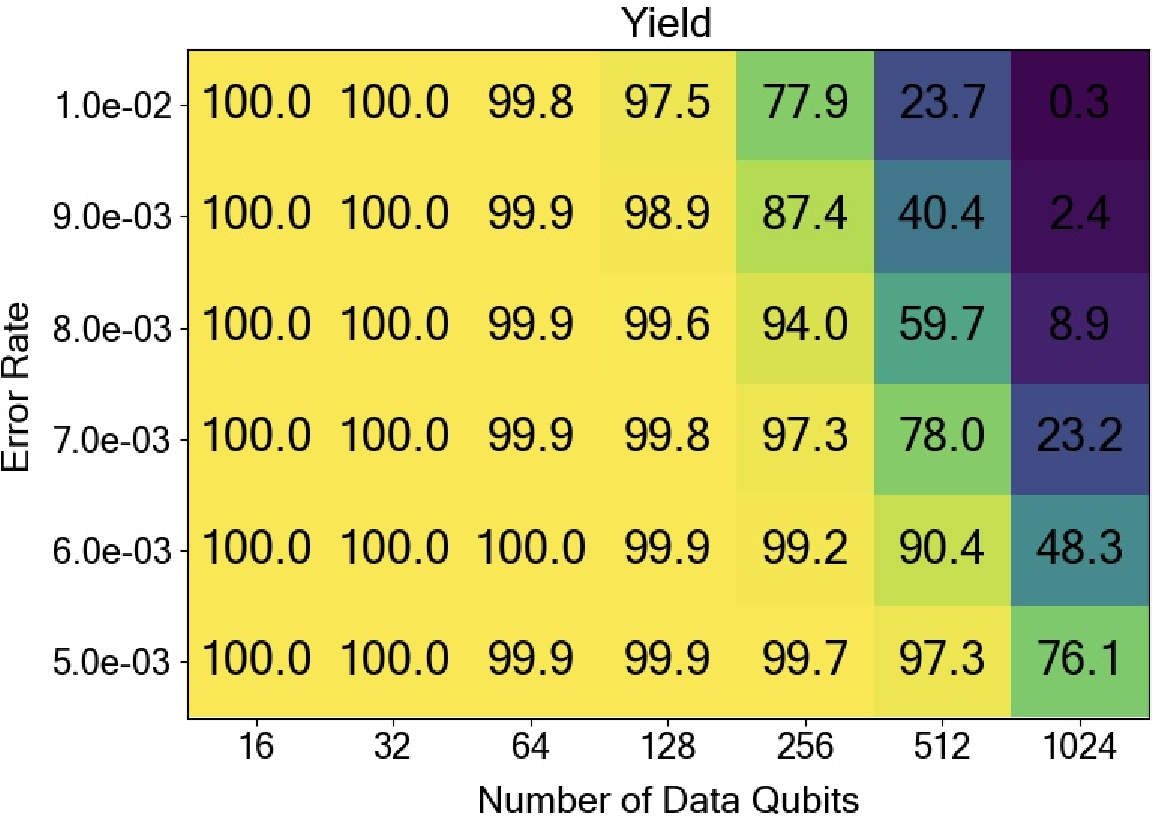}
         \caption{Four redundant qubits}
         \label{fig:five_Yield_rr4_v4}
     \end{subfigure}
     \begin{subfigure}[b]{0.334\textwidth}
         \centering
         \captionsetup{justification=centering}
         \includegraphics[width=\textwidth]{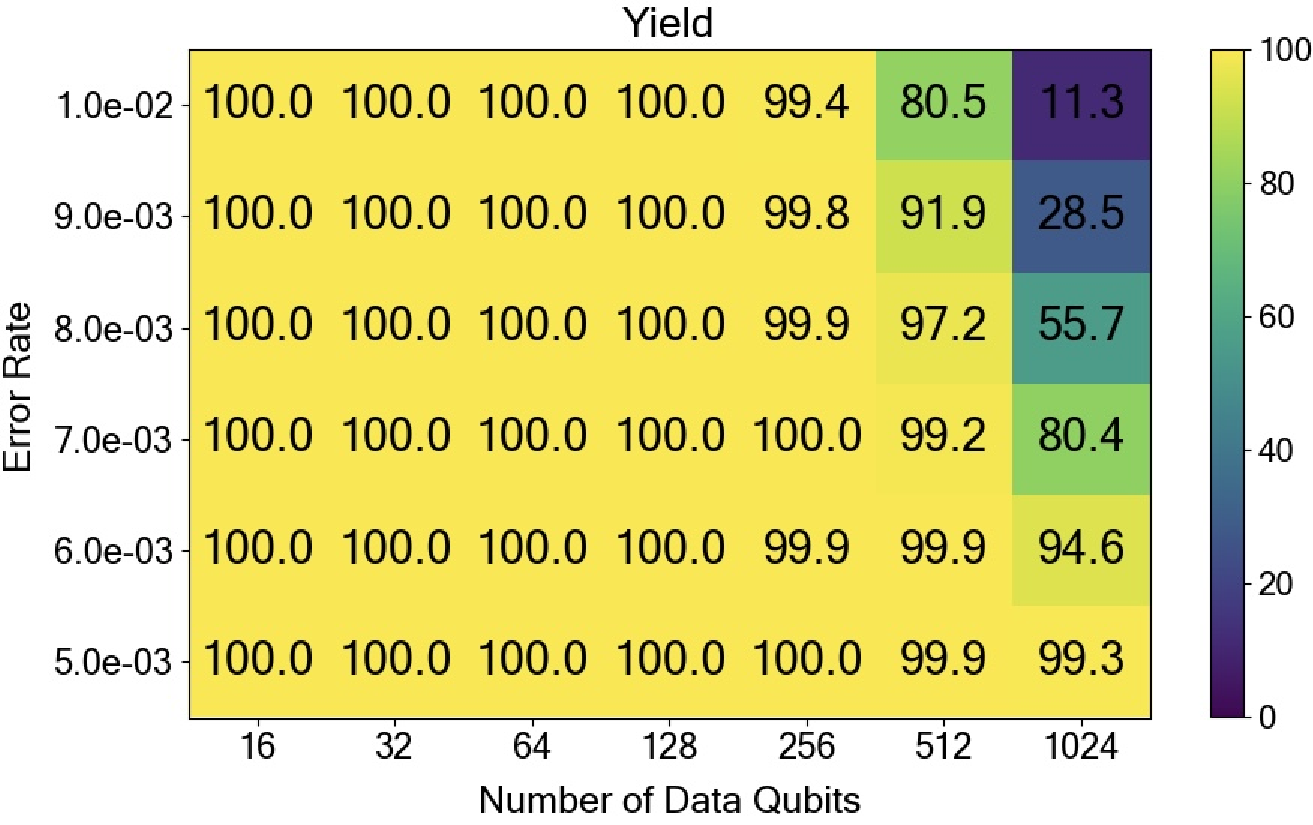}
         \caption{Eight redundant qubits}
         \label{fig:five_Yield_rr8_v6}
     \end{subfigure}
    
        \caption{qRAM yields according to manufacturing error rates, numbers of logical qubits, and numbers of redundant logical qubits. The yield is higher for colors closer to yellow and green, while the yield is lower for colors closer to blue and indigo. From (a) to (e), the number of redundant qubits is sequentially set as 0, 1, 2, 4, and 8, and the surface code distance of the logical qubits of all qRAMs is set to 3.}
        \label{fig:yield}
\end{figure*}

Figure~\ref{fig:yield}, we also compared the yield of qRAM by varying experimental parameters, namely, the fabrication error rate, number of logical qubits in the qRAM, and number of redundant logical qubits. For all five sub-figures, the x-axis represents the number of logical qubits, and the y-axis represents different degrees of fabrication error rate from 0.5\% to 1\%. To make it easier to distinguish visually, we use the colormap representation. The higher the yield, the closer to the yellow color, and the lower the yield, the closer to the indigo color. In all experiments, each logical qubit was encoded with QEC 3.

As shown in the figure, increasing the number of redundant qubits increased the yields of qRAMs with the same fabrication error rate and number of logical qubits. The same phenomenon was observed in qRAMs with small and large numbers of logical qubits. In the qRAM with 16 memory cells and no redundant qubits under a 1\% fabrication error rate, the yield was 82.05\%. In the same qRAM, one redundant qubit improved the yield to 98.18\% (an approximate improvement of 16\%). After increasing the number of logical qubits eight times (from 16 to 128) without changing the fabrication error rate, the yield reduced to 19.94\% with no redundant qubits, but after adding 1, 2, 4, and 8 redundant qubits, the yield improved by 53.35\%, 78.43\%, 97.56\%, and 100\%, respectively. The same tendency was observed after doubling the number of logical qubits from 128 to 256. In the absence of redundant qubits, the yield reached only 4.12\% but after adding 1, 2, 4, and 8 redundant qubits, the yield increased to 17.58\%, 37.94\%, 77.98\%, and 99.42\%, respectively. These results confirm that the qRAM yield can be significantly improved by employing a small number of redundant qubits relative to the number of logical qubits constituting the qRAM.

\subsubsection*{Resource Overhead}
\label{sec:resource_overhead}

The proposed qRAM requires additional physical qubits to support the redundant qubits. Like the QEC scheme, this scheme introduces resource overheads on the number of physical qubits. Therefore, the number of additional physical qubits required in our method must be compared with that of QEC. To consider the resource overhead of our proposed qRAM architecture, we analyzed the total number of physical qubits constructed into the qRAM memory cells and peripherals. The number of physical qubits along the right axis of Figure~\ref{fig:simulation result} refers to the number of qubits required for the memory cells and peripheral parts. 
For qRAMs with the same number of logical qubits, increasing the QEC degree dramatically increased the required number of physical qubits. When eight redundant qubits were added to a qRAM with a given number of logical qubits, the resource overhead was similar to the number of physical qubits required for QEC 3. In other words, the resource overhead (number of required physical qubits) for improving the yield was much lower in the proposed qRAM than in the conventional method of increasing the degree of QEC.

To analyze the resource overhead in more detail, we determined the numbers of physical qubits required for composing the memory cells and peripheral parts. The factor determining the number of physical qubits for a memory cell depends on the code distance. In terms of QEC degree, the number of required physical qubits $N_{mem}$ is given by
\begin{equation}
    N_{mem} = {d \times (N+X)},
    \label{eq:num_mem_pq}
\end{equation}
where $d$, $N$, and $X$ are the degree of QEC, number of original memory cells, and number of spare memory qubits, respectively.

\begin{equation}
\label{equ:num_peri_pq}
    N_{peri}= 
    \begin{cases}
        3\log_{2}N+N+4, & \text{if } X \le 1+\log_{2}N \\
        2\log_{2}N+N+X+3, & \text{if } X > 1+\log_{2}N
    \end{cases}
\end{equation}
\begin{table}[!ht]
\centering
\setlength{\tabcolsep}{1.8pt}    
\renewcommand{\arraystretch}{1.3}
\label{tab:evaluation_result}
\resizebox{\linewidth}{!}{
\begin{tabular}{|c|c|c|c|c|c|c|c|c|c|c|c|c|c|c|c|c|c|c|c|} 
\hline
\multirow{3}{*}{\begin{tabular}[c]{@{}c@{}}\textbf{Code}\\\textbf{ Distance}\end{tabular}} & \multirow{3}{*}{\begin{tabular}[c]{@{}c@{}}\textbf{Number of}\\\textbf{ Logical }\\\textbf{Qubits}\end{tabular}} & \multicolumn{10}{c|}{\textbf{Number of Physical Qubits}}                                                                                                                               & \multicolumn{8}{c|}{\textbf{Resource Overhead (\%)}}                                                                                                                                                                                                                                                                                                                                                                           \\ 
\cline{3-20}
                                                                                           &                                                                                                                  & \multicolumn{2}{c|}{\textbf{Original}} & \multicolumn{2}{c|}{\textbf{RR1}} & \multicolumn{2}{c|}{\textbf{RR2}} & \multicolumn{2}{c|}{\textbf{RR4}} & \multicolumn{2}{c|}{\textbf{RR8}} & \multicolumn{2}{c|}{\textbf{RR1}}                                                                     & \multicolumn{2}{c|}{\textbf{RR2}}                                                                     & \multicolumn{2}{c|}{\textbf{RR4}}                                                                     & \multicolumn{2}{c|}{\textbf{RR8}}                                                                      \\ 
\cline{3-20}
                                                                                           &                                                                                                                  & \textbf{Mem} & \textbf{Peri}           & \textbf{Mem} & \textbf{Peri}      & \textbf{Mem} & \textbf{Peri}      & \textbf{Mem} & \textbf{Peri}      & \textbf{Mem} & \textbf{Peri}      & \textbf{Mem}                                      & \textbf{Peri}                                     & \textbf{Mem}                                      & \textbf{Peri}                                     & \textbf{Mem}                                      & \textbf{Peri}                                     & \textbf{Mem}                                      & \textbf{Peri}                                      \\ 
\hhline{|====================|}
\multirow{7}{*}{\textbf{QEC 3}}                                                            & 16                                                                                                               & 272          & 374                     & 289          & 544                & 306          & 544                & 340          & 544                & 408          & 595                & {\cellcolor[rgb]{0.749,0.749,0.749}}6.25          & {\cellcolor[rgb]{0.749,0.749,0.749}}45.45         & {\cellcolor[rgb]{0.749,0.749,0.749}}12.5          & {\cellcolor[rgb]{0.749,0.749,0.749}}45.5          & {\cellcolor[rgb]{0.749,0.749,0.749}}25.0          & {\cellcolor[rgb]{0.749,0.749,0.749}}45.5          & {\cellcolor[rgb]{0.749,0.749,0.749}}50.0          & {\cellcolor[rgb]{0.749,0.749,0.749}}59.1           \\
                                                                                           & 32                                                                                                               & 544          & 663                     & 561          & 867                & 578          & 867                & 612          & 867                & 680          & 901                & {\cellcolor[rgb]{0.749,0.749,0.749}}3.13          & {\cellcolor[rgb]{0.749,0.749,0.749}}30.77         & {\cellcolor[rgb]{0.749,0.749,0.749}}6.25          & {\cellcolor[rgb]{0.749,0.749,0.749}}30.77         & {\cellcolor[rgb]{0.749,0.749,0.749}}12.5          & {\cellcolor[rgb]{0.749,0.749,0.749}}30.8          & {\cellcolor[rgb]{0.749,0.749,0.749}}25.0          & {\cellcolor[rgb]{0.749,0.749,0.749}}35.9           \\
                                                                                           & 64                                                                                                               & 1,088        & 1,224                   & 1,105        & 1,462              & 1,122        & 1,462              & 1,156        & 1,462              & 1,224        & 1,479              & {\cellcolor[rgb]{0.749,0.749,0.749}}1.56          & {\cellcolor[rgb]{0.749,0.749,0.749}}19.44         & {\cellcolor[rgb]{0.749,0.749,0.749}}3.13          & {\cellcolor[rgb]{0.749,0.749,0.749}}19.44         & {\cellcolor[rgb]{0.749,0.749,0.749}}6.25          & {\cellcolor[rgb]{0.749,0.749,0.749}}19.44         & {\cellcolor[rgb]{0.749,0.749,0.749}}12.5          & {\cellcolor[rgb]{0.749,0.749,0.749}}20.8           \\
                                                                                           & 128                                                                                                              & 2,176        & 2,329                   & 2,193        & 2,601              & 2,210        & 2,601              & 2,244        & 2,601              & 2,312        & 2,601              & {\cellcolor[rgb]{0.749,0.749,0.749}}0.78          & {\cellcolor[rgb]{0.749,0.749,0.749}}11.68         & {\cellcolor[rgb]{0.749,0.749,0.749}}1.56          & {\cellcolor[rgb]{0.749,0.749,0.749}}11.68         & {\cellcolor[rgb]{0.749,0.749,0.749}}3.13          & {\cellcolor[rgb]{0.749,0.749,0.749}}11.68         & {\cellcolor[rgb]{0.749,0.749,0.749}}6.25          & {\cellcolor[rgb]{0.749,0.749,0.749}}11.68          \\
                                                                                           & 256                                                                                                              & 4,352        & 4,522                   & 4,369        & 4,828              & 4,386        & 4,828              & 4,420        & 4,828              & 4,488        & 4,828              & {\cellcolor[rgb]{0.749,0.749,0.749}}0.39          & {\cellcolor[rgb]{0.749,0.749,0.749}}6.77          & {\cellcolor[rgb]{0.749,0.749,0.749}}0.78          & {\cellcolor[rgb]{0.749,0.749,0.749}}6.77          & {\cellcolor[rgb]{0.749,0.749,0.749}}1.56          & {\cellcolor[rgb]{0.749,0.749,0.749}}6.77          & {\cellcolor[rgb]{0.749,0.749,0.749}}3.13          & {\cellcolor[rgb]{0.749,0.749,0.749}}6.77           \\
                                                                                           & 512                                                                                                              & 8,704        & 8,891                   & 8,721        & 9,231              & 8,738        & 9,231              & 8,772        & 9,231              & 8,840        & 9,231              & {\cellcolor[rgb]{0.749,0.749,0.749}}0.20          & {\cellcolor[rgb]{0.749,0.749,0.749}}3.82          & {\cellcolor[rgb]{0.749,0.749,0.749}}0.39          & {\cellcolor[rgb]{0.749,0.749,0.749}}3.82          & {\cellcolor[rgb]{0.749,0.749,0.749}}0.78          & {\cellcolor[rgb]{0.749,0.749,0.749}}3.82          & {\cellcolor[rgb]{0.749,0.749,0.749}}1.56          & {\cellcolor[rgb]{0.749,0.749,0.749}}3.82           \\
                                                                                           & \textbf{1,024}                                                                                                   & 17,408       & 17,612                  & 17,425       & 17,986             & 17,442       & 17,986             & 17,476       & 17,986             & 17,544       & 17,986             & {\cellcolor[rgb]{0.749,0.749,0.749}}\textbf{0.10} & {\cellcolor[rgb]{0.749,0.749,0.749}}\textbf{2.12} & {\cellcolor[rgb]{0.749,0.749,0.749}}\textbf{0.20} & {\cellcolor[rgb]{0.749,0.749,0.749}}\textbf{2.12} & {\cellcolor[rgb]{0.749,0.749,0.749}}\textbf{0.39} & {\cellcolor[rgb]{0.749,0.749,0.749}}\textbf{2.12} & {\cellcolor[rgb]{0.749,0.749,0.749}}\textbf{0.78} & {\cellcolor[rgb]{0.749,0.749,0.749}}\textbf{2.12}  \\ 
\hhline{|====================|}
\multirow{7}{*}{\textbf{QEC 5}}                                                            & 16                                                                                                               & 784          & 1,078                   & 833          & 1,568              & 882          & 1,568              & 980          & 1,568              & 1,176        & 1,715              & {\cellcolor[rgb]{0.749,0.749,0.749}}6.25          & {\cellcolor[rgb]{0.749,0.749,0.749}}45.45         & {\cellcolor[rgb]{0.749,0.749,0.749}}12.5          & {\cellcolor[rgb]{0.749,0.749,0.749}}45.5          & {\cellcolor[rgb]{0.749,0.749,0.749}}25.0          & {\cellcolor[rgb]{0.749,0.749,0.749}}45.5          & {\cellcolor[rgb]{0.749,0.749,0.749}}50.0          & {\cellcolor[rgb]{0.749,0.749,0.749}}59.1           \\
                                                                                           & 32                                                                                                               & 1,568        & 1,911                   & 1,617        & 2,499              & 1,666        & 2,499              & 1,764        & 2,499              & 1,960        & 2,597              & {\cellcolor[rgb]{0.749,0.749,0.749}}3.13          & {\cellcolor[rgb]{0.749,0.749,0.749}}30.77         & {\cellcolor[rgb]{0.749,0.749,0.749}}6.25          & {\cellcolor[rgb]{0.749,0.749,0.749}}30.77         & {\cellcolor[rgb]{0.749,0.749,0.749}}12.5          & {\cellcolor[rgb]{0.749,0.749,0.749}}30.8          & {\cellcolor[rgb]{0.749,0.749,0.749}}25.0          & {\cellcolor[rgb]{0.749,0.749,0.749}}35.9           \\
                                                                                           & 64                                                                                                               & 3,136        & 3,528                   & 3,185        & 4,214              & 3,234        & 4,214              & 3,332        & 4,214              & 3,528        & 4,263              & {\cellcolor[rgb]{0.749,0.749,0.749}}1.56          & {\cellcolor[rgb]{0.749,0.749,0.749}}19.44         & {\cellcolor[rgb]{0.749,0.749,0.749}}3.13          & {\cellcolor[rgb]{0.749,0.749,0.749}}19.44         & {\cellcolor[rgb]{0.749,0.749,0.749}}6.25          & {\cellcolor[rgb]{0.749,0.749,0.749}}19.44         & {\cellcolor[rgb]{0.749,0.749,0.749}}12.5          & {\cellcolor[rgb]{0.749,0.749,0.749}}20.8           \\
                                                                                           & 128                                                                                                              & 6,272        & 6,713                   & 6,321        & 7,497              & 6,370        & 7,497              & 6,468        & 7,497              & 6,664        & 7,497              & {\cellcolor[rgb]{0.749,0.749,0.749}}0.78          & {\cellcolor[rgb]{0.749,0.749,0.749}}11.68         & {\cellcolor[rgb]{0.749,0.749,0.749}}1.56          & {\cellcolor[rgb]{0.749,0.749,0.749}}11.68         & {\cellcolor[rgb]{0.749,0.749,0.749}}3.13          & {\cellcolor[rgb]{0.749,0.749,0.749}}11.68         & {\cellcolor[rgb]{0.749,0.749,0.749}}6.25          & {\cellcolor[rgb]{0.749,0.749,0.749}}11.68          \\
                                                                                           & 256                                                                                                              & 12,544       & 13,034                  & 12,593       & 13,916             & 12,642       & 13,916             & 12,740       & 13,916             & 12,936       & 13,916             & {\cellcolor[rgb]{0.749,0.749,0.749}}0.39          & {\cellcolor[rgb]{0.749,0.749,0.749}}6.77          & {\cellcolor[rgb]{0.749,0.749,0.749}}0.78          & {\cellcolor[rgb]{0.749,0.749,0.749}}6.77          & {\cellcolor[rgb]{0.749,0.749,0.749}}1.56          & {\cellcolor[rgb]{0.749,0.749,0.749}}6.77          & {\cellcolor[rgb]{0.749,0.749,0.749}}3.13          & {\cellcolor[rgb]{0.749,0.749,0.749}}6.77           \\
                                                                                           & 512                                                                                                              & 25,088       & 25,627                  & 25,137       & 26,607             & 25,186       & 26,607             & 25,284       & 26,607             & 25,480       & 26,607             & {\cellcolor[rgb]{0.749,0.749,0.749}}0.20          & {\cellcolor[rgb]{0.749,0.749,0.749}}3.82          & {\cellcolor[rgb]{0.749,0.749,0.749}}0.39          & {\cellcolor[rgb]{0.749,0.749,0.749}}3.82          & {\cellcolor[rgb]{0.749,0.749,0.749}}0.78          & {\cellcolor[rgb]{0.749,0.749,0.749}}3.82          & {\cellcolor[rgb]{0.749,0.749,0.749}}1.56          & {\cellcolor[rgb]{0.749,0.749,0.749}}3.82           \\
                                                                                           & 1,024                                                                                                            & 50,176       & 50,764                  & 50,225       & 51,842             & 50,274       & 51,842             & 50,372       & 51,842             & 50,568       & 51,842             & {\cellcolor[rgb]{0.749,0.749,0.749}}0.10          & {\cellcolor[rgb]{0.749,0.749,0.749}}2.12          & {\cellcolor[rgb]{0.749,0.749,0.749}}0.20          & {\cellcolor[rgb]{0.749,0.749,0.749}}2.12          & {\cellcolor[rgb]{0.749,0.749,0.749}}0.39          & {\cellcolor[rgb]{0.749,0.749,0.749}}2.12          & {\cellcolor[rgb]{0.749,0.749,0.749}}0.78          & {\cellcolor[rgb]{0.749,0.749,0.749}}2.12           \\ 
\hhline{|====================|}
\multirow{7}{*}{\textbf{QEC 7}}                                                            & 16                                                                                                               & 1,552        & 2,134                   & 1,649        & 3,104              & 1,746        & 3,104              & 1,940        & 3,104              & 2,328        & 3,395              & {\cellcolor[rgb]{0.749,0.749,0.749}}6.25          & {\cellcolor[rgb]{0.749,0.749,0.749}}45.45         & {\cellcolor[rgb]{0.749,0.749,0.749}}12.5          & {\cellcolor[rgb]{0.749,0.749,0.749}}45.5          & {\cellcolor[rgb]{0.749,0.749,0.749}}25.0          & {\cellcolor[rgb]{0.749,0.749,0.749}}45.5          & {\cellcolor[rgb]{0.749,0.749,0.749}}50.0          & {\cellcolor[rgb]{0.749,0.749,0.749}}59.1           \\
                                                                                           & 32                                                                                                               & 3,104        & 3,783                   & 3,201        & 4,947              & 3,298        & 4,947              & 3,492        & 4,947              & 3,880        & 5,141              & {\cellcolor[rgb]{0.749,0.749,0.749}}3.13          & {\cellcolor[rgb]{0.749,0.749,0.749}}30.77         & {\cellcolor[rgb]{0.749,0.749,0.749}}6.25          & {\cellcolor[rgb]{0.749,0.749,0.749}}30.77         & {\cellcolor[rgb]{0.749,0.749,0.749}}12.5          & {\cellcolor[rgb]{0.749,0.749,0.749}}30.8          & {\cellcolor[rgb]{0.749,0.749,0.749}}25.0          & {\cellcolor[rgb]{0.749,0.749,0.749}}35.9           \\
                                                                                           & 64                                                                                                               & 6,208        & 6,984                   & 6,305        & 8,342              & 6,402        & 8,342              & 6,596        & 8,342              & 6,984        & 8,439              & {\cellcolor[rgb]{0.749,0.749,0.749}}1.56          & {\cellcolor[rgb]{0.749,0.749,0.749}}19.44         & {\cellcolor[rgb]{0.749,0.749,0.749}}3.13          & {\cellcolor[rgb]{0.749,0.749,0.749}}19.44         & {\cellcolor[rgb]{0.749,0.749,0.749}}6.25          & {\cellcolor[rgb]{0.749,0.749,0.749}}19.44         & {\cellcolor[rgb]{0.749,0.749,0.749}}12.5          & {\cellcolor[rgb]{0.749,0.749,0.749}}20.8           \\
                                                                                           & 128                                                                                                              & 12,416       & 13,289                  & 12,513       & 14,841             & 12,610       & 14,841             & 12,804       & 14,841             & 13,192       & 14,841             & {\cellcolor[rgb]{0.749,0.749,0.749}}0.78          & {\cellcolor[rgb]{0.749,0.749,0.749}}11.68         & {\cellcolor[rgb]{0.749,0.749,0.749}}1.56          & {\cellcolor[rgb]{0.749,0.749,0.749}}11.68         & {\cellcolor[rgb]{0.749,0.749,0.749}}3.13          & {\cellcolor[rgb]{0.749,0.749,0.749}}11.68         & {\cellcolor[rgb]{0.749,0.749,0.749}}6.25          & {\cellcolor[rgb]{0.749,0.749,0.749}}11.68          \\
                                                                                           & 256                                                                                                              & 24,832       & 25,802                  & 24,929       & 27,548             & 25,026       & 27,548             & 25,220       & 27,548             & 25,608       & 27,548             & {\cellcolor[rgb]{0.749,0.749,0.749}}0.39          & {\cellcolor[rgb]{0.749,0.749,0.749}}6.77          & {\cellcolor[rgb]{0.749,0.749,0.749}}0.78          & {\cellcolor[rgb]{0.749,0.749,0.749}}6.77          & {\cellcolor[rgb]{0.749,0.749,0.749}}1.56          & {\cellcolor[rgb]{0.749,0.749,0.749}}6.77          & {\cellcolor[rgb]{0.749,0.749,0.749}}3.13          & {\cellcolor[rgb]{0.749,0.749,0.749}}6.77           \\
                                                                                           & 512                                                                                                              & 49,664       & 50,731                  & 49,761       & 52,671             & 49,858       & 52,671             & 50,052       & 52,671             & 50,440       & 52,671             & {\cellcolor[rgb]{0.749,0.749,0.749}}0.20          & {\cellcolor[rgb]{0.749,0.749,0.749}}3.82          & {\cellcolor[rgb]{0.749,0.749,0.749}}0.39          & {\cellcolor[rgb]{0.749,0.749,0.749}}3.82          & {\cellcolor[rgb]{0.749,0.749,0.749}}0.78          & {\cellcolor[rgb]{0.749,0.749,0.749}}3.82          & {\cellcolor[rgb]{0.749,0.749,0.749}}1.56          & {\cellcolor[rgb]{0.749,0.749,0.749}}3.82           \\
                                                                                           & 1,024                                                                                                            & 99,328       & 100,492                 & 99,425       & 102,626            & 99,522       & 102,626            & 99,716       & 102,626            & 100,104      & 102,626            & {\cellcolor[rgb]{0.749,0.749,0.749}}0.10          & {\cellcolor[rgb]{0.749,0.749,0.749}}2.12          & {\cellcolor[rgb]{0.749,0.749,0.749}}0.20          & {\cellcolor[rgb]{0.749,0.749,0.749}}2.12          & {\cellcolor[rgb]{0.749,0.749,0.749}}0.39          & {\cellcolor[rgb]{0.749,0.749,0.749}}2.12          & {\cellcolor[rgb]{0.749,0.749,0.749}}0.78          & {\cellcolor[rgb]{0.749,0.749,0.749}}2.12           \\ 
\hhline{|====================|}
\multirow{7}{*}{\textbf{QEC 9}}                                                            & 16                                                                                                               & 2,576        & 3,542                   & 2,737        & 5,152              & 2,898        & 5,152              & 3,220        & 5,152              & 3,864        & 5,635              & {\cellcolor[rgb]{0.749,0.749,0.749}}6.25          & {\cellcolor[rgb]{0.749,0.749,0.749}}45.45         & {\cellcolor[rgb]{0.749,0.749,0.749}}12.5          & {\cellcolor[rgb]{0.749,0.749,0.749}}45.5          & {\cellcolor[rgb]{0.749,0.749,0.749}}25.0          & {\cellcolor[rgb]{0.749,0.749,0.749}}45.5          & {\cellcolor[rgb]{0.749,0.749,0.749}}50.0          & {\cellcolor[rgb]{0.749,0.749,0.749}}59.1           \\
                                                                                           & 32                                                                                                               & 5,152        & 6,279                   & 5,313        & 8,211              & 5,474        & 8,211              & 5,796        & 8,211              & 6,440        & 8,533              & {\cellcolor[rgb]{0.749,0.749,0.749}}3.13          & {\cellcolor[rgb]{0.749,0.749,0.749}}30.77         & {\cellcolor[rgb]{0.749,0.749,0.749}}6.25          & {\cellcolor[rgb]{0.749,0.749,0.749}}30.77         & {\cellcolor[rgb]{0.749,0.749,0.749}}12.5          & {\cellcolor[rgb]{0.749,0.749,0.749}}30.8          & {\cellcolor[rgb]{0.749,0.749,0.749}}25.0          & {\cellcolor[rgb]{0.749,0.749,0.749}}35.9           \\
                                                                                           & 64                                                                                                               & 10,304       & 11,592                  & 10,465       & 13,846             & 10,626       & 13,846             & 10,948       & 13,846             & 11,592       & 14,007             & {\cellcolor[rgb]{0.749,0.749,0.749}}1.56          & {\cellcolor[rgb]{0.749,0.749,0.749}}19.44         & {\cellcolor[rgb]{0.749,0.749,0.749}}3.13          & {\cellcolor[rgb]{0.749,0.749,0.749}}19.44         & {\cellcolor[rgb]{0.749,0.749,0.749}}19.44         & {\cellcolor[rgb]{0.749,0.749,0.749}}12.66         & {\cellcolor[rgb]{0.749,0.749,0.749}}12.5          & {\cellcolor[rgb]{0.749,0.749,0.749}}20.8           \\
                                                                                           & 128                                                                                                              & 20,608       & 22,057                  & 20,769       & 24,633             & 20,930       & 24,633             & 21,252       & 24,633             & 21,896       & 24,633             & {\cellcolor[rgb]{0.749,0.749,0.749}}0.78          & {\cellcolor[rgb]{0.749,0.749,0.749}}11.68         & {\cellcolor[rgb]{0.749,0.749,0.749}}1.56          & {\cellcolor[rgb]{0.749,0.749,0.749}}11.68         & {\cellcolor[rgb]{0.749,0.749,0.749}}3.13          & {\cellcolor[rgb]{0.749,0.749,0.749}}11.68         & {\cellcolor[rgb]{0.749,0.749,0.749}}6.25          & {\cellcolor[rgb]{0.749,0.749,0.749}}11.68          \\
                                                                                           & 256                                                                                                              & 41,216       & 42,826                  & 41,377       & 45,724             & 41,538       & 45,724             & 41,860       & 45,724             & 42,504       & 45,724             & {\cellcolor[rgb]{0.749,0.749,0.749}}0.39          & {\cellcolor[rgb]{0.749,0.749,0.749}}6.77          & {\cellcolor[rgb]{0.749,0.749,0.749}}0.78          & {\cellcolor[rgb]{0.749,0.749,0.749}}6.77          & {\cellcolor[rgb]{0.749,0.749,0.749}}1.56          & {\cellcolor[rgb]{0.749,0.749,0.749}}6.77          & {\cellcolor[rgb]{0.749,0.749,0.749}}3.13          & {\cellcolor[rgb]{0.749,0.749,0.749}}6.77           \\
                                                                                           & 512                                                                                                              & 82,432       & 84,203                  & 82,593       & 87,423             & 82,754       & 87,423             & 83,076       & 87,423             & 83,720       & 87,423             & {\cellcolor[rgb]{0.749,0.749,0.749}}0.20          & {\cellcolor[rgb]{0.749,0.749,0.749}}3.82          & {\cellcolor[rgb]{0.749,0.749,0.749}}0.39          & {\cellcolor[rgb]{0.749,0.749,0.749}}3.82          & {\cellcolor[rgb]{0.749,0.749,0.749}}0.78          & {\cellcolor[rgb]{0.749,0.749,0.749}}3.82          & {\cellcolor[rgb]{0.749,0.749,0.749}}1.56          & {\cellcolor[rgb]{0.749,0.749,0.749}}3.82           \\
                                                                                           & 1,024                                                                                                            & 164,864      & 166,796                 & 165,025      & 170,338            & 165,186      & 170,338            & 165,508      & 170,338            & 166,152      & 170,338            & {\cellcolor[rgb]{0.749,0.749,0.749}}0.10          & {\cellcolor[rgb]{0.749,0.749,0.749}}2.12          & {\cellcolor[rgb]{0.749,0.749,0.749}}0.20          & {\cellcolor[rgb]{0.749,0.749,0.749}}2.12          & {\cellcolor[rgb]{0.749,0.749,0.749}}0.39          & {\cellcolor[rgb]{0.749,0.749,0.749}}2.12          & {\cellcolor[rgb]{0.749,0.749,0.749}}0.78          & {\cellcolor[rgb]{0.749,0.749,0.749}}2.12           \\
\hline
\end{tabular}}
    \captionof{table}{Simulated numbers of physical (memory and periphery) qubits and resource overheads (\%) in the absence of a repair scheme (Original) and with 1, 2, 4, 8 redundant repair qubits (RR1, RR2, RR4, and RR8) respectively.}
    \label{tab:evaluation_result}
\end{table}

Meanwhile, the required number of physical qubits for the peripheral part is based on a quantum circuit model. The peripheral part can be largely divided into three parts such as address qubits, routing nodes, and read/write part. Using address qubits, we first identified input and spare addresses. For example, suppose that there are $N$ original memory qubits and $X$ spare memory qubits to replace $X$ faults. To represent the $N$ original memory qubits, the input address requires $\log_{2}(N)$ qubits. A spare address requires an additional $\log_{2}(N)$ qubits because an arbitrary memory cell in qRAM must be addressed within the same time. Additionally, the routing nodes are the qubits needed for addressing, which require $RFQ$ and the $|1\rangle$ qubits to support the redundant repair scheme. The proposed qRAM also requires original and spare memory qubits to support the address routing circuit, which is a binary tree. Addressing the original memory qubits requires $N-1$ nodes for a complete binary tree. In contrast, the required number of nodes for addressing the spare memory qubits depends on the number of physically implemented spare qubits ($X$). This is because the spare routing tree only needs to be large enough to uniquely address these $X$ spare cells, not the full $N$ ($=2^n$) address space. Therefore, the number of nodes in the spare tree is determined by this quantity $X$ (where $X \ll N$), making it significantly smaller than the $N-1$ nodes required for the original tree. $DQ$, $Readout$, and $R/W$ qubits are required for memory read and write. Therefore, the number of physical qubits for peripheral part $N_{peri}$ in the circuit model is calculated as equation~\eqref{equ:num_peri_pq}.

Table~\ref{tab:evaluation_result} shows the detailed resource overheads obtained in all experiments. This table compares the numbers of physical qubits required for the memory cell and peripherals in qRAMs with different degrees of QEC, numbers of logical qubits, and numbers of redundant qubits. When no redundant repair scheme was applied, the overhead was compared in terms of the number of physical qubits for different numbers of redundant qubits (gray cells in the table). For example, in the qRAM with 1,024 logical qubits, the number of additional physical qubits required to support eight redundant qubits incurred memory and peripheral overheads of 0.78\% and 2.12\%, respectively, relative to the total number of physical qubits. In numerical terms, the requirement is 35,530 physical qubits, versus 35,020 qubits in the absence of repair application. Moreover, even if the degree of QEC in a qRAM increases up to QEC 9 with the same number of logical qubits, only 1.45\% of 331,660 physical qubits, i.e., 4,830 additional qubits were used for the redundant repair scheme. Thus, even with a high degree of QEC, we confirmed the resource overhead of our scheme is still low. Still, it is worth noting the increment of resource overhead with the higher degree of QEC. As the logical qubits of qRAM and redundant qubits use the same code distance, a higher degree of QEC will increase resource overhead for redundant qubits as well.

\section*{Related Works}
\label{sec:related_works}

Practical implementations of universal quantum computing require the mitigation of quantum errors. Recent studies~\cite{qin2022ao, strikis2023, vovrosh2021sm, tang2016rs} have focused on suppressing the two well-known noise channels, i.e., the depolarizing error and fabrication defect. 

Vovrosh et al.~\cite{vovrosh2021sm} proposed a simple but effective error-mitigation technique for depolarizing error channels. As an error model ansatz, they assumed a deep quantum circuit with global depolarizing error channels. They extracted the error-free results from the noisy data in this error model. Error mitigation was first demonstrated in entanglement measurements and then in real-time dynamics of confinement in quantum spin chains. Their error-mitigation technique was deemed applicable to broader settings and in numerical simulations of more general tasks using a realistic error model. Despite its advantages, this protocol incurs a high computational overhead because the system size grows exponentially with the number of randomized unitaries and random measurements. This work is designed to mitigate dynamic noise errors that occur during the execution of the algorithm. In contrast, our work focuses on solving the problem of static fabrication defects, which are permanent, physical flaws in the qubits. Our proposed BISR mechanism is an architectural solution to bypass these static defects, a problem fundamentally distinct from mitigating dynamic runtime noise.

Tang et al.~\cite{tang2016rs} proposed a defect-tolerant surface code topology with resistance to spare fabrication defects, which can be practically implemented on universal quantum computing. The authors stated that when a disk is folded into $N$ layers, a defective physical qubit can be replaced by a working physical qubit on the same unit cell from a different layer. This technique uses one additional layer in the real physical qubit topology to store the spare fabrication errors that replace the defective physical qubits. Tang et al. also demonstrated the robustness of the surface code topology by fine-tuning the operation of the logical measure qubits. When coupling two physical qubits into one logical measure qubit, they bridged the two layers of the physical units via two schemes based on superconducting flux qubits and Xmon qubits on the circuit level. Although this technique effectively replaces a defective physical qubit, the additional physical topology incurs a large overhead. The fundamental difference from our work lies in the solution level. Tang's approach is a physical level topology modification, requiring complex multi-layer fabrication and design to replace a defect. In contrast, our work proposes an architectural level solution that assumes a standard physical layout. We introduce a logical built-in self-repair mechanism, analogous to classical RAM repair, which logically bypasses defects at the address routing stage. This architectural approach avoids the significant manufacturing and design overhead of modifying the physical qubit topology and offers a more flexible, scalable repair scheme based on pre-identified defect information.

Finally, the robustness of qRAM has recently been investigated. Giovannetti et al.~\cite{giovannetti2008qr} developed the Bucket Brigade structure, which improves the error robustness by reducing the number of gate activations in memory addressing from those of the Fanout structure. Since then, further advancements have focused on optimizing query parallelization and estimating resources for fault-tolerant implementations~\cite{paler2020pt, di2020ft, hann2021ro}. Most recently, the scope of robustness has extended to handling fabrication defects within the routing hardware; for instance, the Faulty Towers architecture~\cite{weiss_faulty_2024} proposes a strategy to recover operational routing paths by identifying and bypassing defective routers. However, these studies, including the router-recovery schemes, strictly focus on errors and defects within the memory addressing mechanism. They aim to ensure that queries are correctly routed to the target address, but do not guarantee the integrity of the memory cells themselves. \textit{To the best of our knowledge}, our work is the first to address fabrication defects in the memory cells used for read and write operations, providing a necessary complement to existing routing-focused solutions.

\section*{Discussion}
\label{sec:Discussion}
In this study, we propose a redundancy repair technique that maximizes QRAM yield by mitigating defects in memory cells. To advance the realization of a truly comprehensive fault-tolerant QRAM architecture, it is crucial to establish the relationship between recent advancements in router defect mitigation, particularly the Faulty Towers architecture~\cite{weiss_faulty_2024}, and our own research. The fundamental difference between the two approaches lies in their primary target of mitigation and their respective methodologies. The Faulty Towers architecture primarily addresses defective routing nodes within the Bucket Brigade structure, employing logical re-routing strategies, such as the IterativeRepair algorithm, to bypass broken connections and restore functionality. In contrast, our study focuses on the memory cells (leaf nodes) where quantum information is stored, replacing defective logical qubits with spare qubits through physical redundancy. Essentially, while Faulty Towers ensures the integrity of the path, our research guarantees the integrity of the destination.

This distinction suggests that the two approaches are complementary, and their integration is essential for handling complex error scenarios. Applying only one method in isolation would leave the system vulnerable to specific types of failures. For example, even if the IterativeRepair algorithm successfully bypasses a faulty router and establishes a valid path, the operation would inevitably fail if the target memory cell at the end of that path (or the re-routed leaf node) is itself defective. Conversely, a functionally intact spare memory cell would be useless if a higher-level router defect prevents any routing path from being established to it. Therefore, a robust QRAM must combine logical re-routing for its tree structure and physical redundancy for its memory cells to survive in situations where defects are distributed across both the router network and memory storage.

Furthermore, a hybrid integration of these strategies offers the most resource-efficient solution to the overall overhead challenge. While each method excels at its primary objective, a comprehensive approach to mitigating manufacturing defects across all components of the QRAM benefits greatly from leveraging its complementary strengths. Our physical redundancy for memory cells, for instance, provides highly efficient mitigation for distributed defects, securing high yield with modest, flexible qubit overhead (e.g., about 1\%). In contrast, logical re-routing methods (Weiss et al.) handle structural defects but involve substantial fixed upfront costs: they necessitate fabricating a physical structure twice the size of the functional memory (e.g., an $n$-bit QRAM for an $(n-1)$-bit unit) and introduce an $O(n^2)$ query latency. These costs become particularly inefficient if re-routing frequently targets unmitigated faulty memory cells. Recognizing these distinct overhead profiles, a hybrid architecture, we contend, is optimal for maximizing efficiency.

A robust QRAM will require a multi-layered approach to defect mitigation. Looking ahead, future research should focus on seamlessly integrating these distinct yet complementary approaches into a truly hybrid fault-tolerant QRAM architecture. By effectively addressing different types of manufacturing defects through distinct methodologies, and by optimally managing resource utilization through their combined strengths, the synergy between path-recovery (Faulty Towers) and destination-recovery (our work) will pave the way for highly reliable and cost-effective, fault-tolerant QRAMs with both high manufacturing yield and operational reliability.

\section*{Conclusion}
\label{sec:conclusion}
qRAM is an essential component of quantum computers and an effective qRAM would realize the full computational benefits of quantum algorithms. Various studies have attempted to improve the fault tolerance of the Bucket Brigade architecture, which is robust against errors because its circuit model uses fewer quantum gates requiring activation than the existing Fanout model. However, while efforts to enhance the fault tolerance of the Bucket Brigade circuitry are ongoing, errors in the qubits of memory cells have been neglected.

Here, we considered the error-proneness of qubits consisting of actual memory cells for fully fault-tolerant qRAM. Additionally, we proposed a redundant repair scheme for qRAM that reduces the number of physical qubits required in the existing QEC method by assigning spare logical qubits. After implementing this scheme, the qRAM yield increased to 99.35\% while the proportion of additional physical qubits was only 1.01\% of all physical qubits. In addition, in qRAMs with 16, 32, 64, 128, 256, 512, and 1,024 logical qubits, the yield improved by 3.05\%, 6.01\%, 12.08\%, 22.09\%, 39.39\%, 62.14\%, and 83.59\%, respectively. Our experimental results confirmed that our redundant repair scheme improves the yield and reduces the resource overheads of qRAM. The proposed scheme will greatly enhance the efficiency of future qRAM involving a large number of mass-produced qubits.

We presented a qRAM architecture for redundancy repair, utilizing superconducting qubit technology, and showed quantum circuit model implementation. Currently, numerous studies are actively exploring various qRAM designs, each based on different qubit technologies. In our approach, the qRAM is specifically designed to accommodate the repair of defective qubits by using redundant qubits. Therefore, the proposed method can be applied regardless of the qubit technologies used for qRAM if redundant qubits are used to replace defective qubits.

\section*{Data availability}
The datasets generated during the current study and the simulation framework are available in the GitHub repository and can be accessed via https://github.com/QCL-PKNU/RR-QRAMSim

\bibliography{manuscript}

\section*{Acknowledgment}
This work was partly supported by Institute for Information \& Communications Technology Planning \& Evaluation (IITP) grant funded by the Korea government (MSIT) [No. 2020-0-00014, A Technology Development of Quantum OS for Fault-tolerant Logical Qubit Computing Environment] and Creation of the quantum information science R\&D ecosystem(based on human resources) through the National Research Foundation of Korea(NRF) funded by the Korean government (Ministry of Science and ICT(MSIT)) (No. RS-2023-00256050).

\section*{Funding}
No funding.

\section*{Author contributions}

D.K. led the majority of this research work. D.K. conceived the redundant repair scheme, designed the overall qRAM architecture, developed the quantum circuit model, implemented the simulation framework, conducted the performance evaluation experiments, performed the yield analysis, and drafted the initial manuscript. S.H. assisted with the experimental setup and contributed to the simulation implementation. S.L. contributed to the theoretical analysis of quantum error correction, assisted with the experimental design, and participated in manuscript revision. Y.H., as the corresponding author, supervised the entire research project, provided guidance on the quantum memory architecture design, secured funding support, and edited the final manuscript. All authors contributed to the discussion of results, reviewed the manuscript, and approved the final version.

\section*{Declarations}
\section*{Competing interests}
The authors declare no competing interests.
\section*{Additional information}

Correspondence and requests for materials should be addressed to Y.H.

\end{document}